\documentclass[10pt,journal,dvipsnames,nofonttune]{IEEEtran}

\usepackage{comment}
\usepackage{xcolor}
\usepackage{listings}
\usepackage{balance}
\usepackage{pgfplots}
\newlength\fheight
\newlength\fwidth
\usepackage{tikzscale}
\pgfplotsset{compat=1.18}
\usepackage{xcolor}
\usepackage{caption}
\usepackage{tikz}
\usepackage{dblfloatfix}
\definecolor{BlueKey}{rgb}{0.1, 0.3, 0.8}  
\usetikzlibrary{calc}
\lstdefinestyle{mystyle-json}{
  commentstyle=\color{gray},
  keywordstyle=\color{BlueKey}, 
  numberstyle=\tiny\color{gray},
  stringstyle=\color{BlueKey},  
  basicstyle=\color{BlueKey}\ttfamily\scriptsize,  
  rulecolor=\color{black},
  breakatwhitespace=true,
  breaklines=true,
  captionpos=b,
  frame=tb,
  numbers=left,
  numbersep=5pt,
  showspaces=false,
  showstringspaces=false,
  showtabs=false,
  tabsize=2,
  xleftmargin=10pt,
  belowskip=-10pt,
}

\lstdefinestyle{mystyle-json-no-numbers}{
  commentstyle=\color{gray},           
  keywordstyle=\color{BlueKey},        
  numberstyle=\tiny\color{gray},       
  stringstyle=\color{BlueKey},      
  basicstyle=\color{BlueKey}\ttfamily\scriptsize, 
  rulecolor=\color{black},
  breakatwhitespace=true,         
  breaklines=true,                 
  captionpos=b,
  frame=tb,
  keepspaces=true,                 
  numbersep=5pt,                  
  showspaces=false,                
  showstringspaces=false,
  showtabs=false,                  
  tabsize=2,
  xleftmargin=10pt,
  belowskip=-10pt,
}

\lstdefinelanguage{json}{
  alsoletter={-},
  keywords={listen,on,off,dst,periodic},
  sensitive=false,
  comment=[l]{\#},
  morecomment=[s]{/*}{*/},
  moredelim=[l][\color{orange}]{\&},
  moredelim=[l][\color{magenta}]{*},
  moredelim=**[il][\color{purple}{:}\color{blue}]{:},   
  morestring=[b]',
  morestring=[b]",
}

\usepackage{tikz}
\usetikzlibrary{patterns}
\usepackage{amsmath}
\usepackage[acronyms,nonumberlist,nopostdot,nomain,nogroupskip,acronymlists={hidden}]{glossaries}
\usepackage{xspace}
\usepackage{float}
\usepackage{enumitem}
\newglossary[algh]{hidden}{acrh}{acnh}{Hidden Acronyms}
\glsdisablehyper

\usepackage{soul}

\usepackage{booktabs}
\usepackage{tabularx}
\usepackage[font=footnotesize]{subcaption}
\usepackage[font=footnotesize]{caption}

\usepackage{url}

\def\framework{AutoRAN\xspace}
\newcommand{\oran}{O-RAN\xspace}
\newcommand{\ran}{\gls{ran}\xspace}

\newcommand*{\cicd}{\gls{ci}/\gls{cd}\xspace}

\newcommand*{\rifc}{RAN-Infrastructure-as-Code\xspace}

\newcommand{\revision}[1]{\textcolor{black}{#1}}

\makeatletter
\def\bstctlcite{\@ifnextchar[{\@bstctlcite}{\@bstctlcite[@auxout]}}
\def\@bstctlcite[#1]#2{\@bsphack
  \@for\@citeb:=#2\do{%
    \edef\@citeb{\expandafter\@firstofone\@citeb}%
    \if@filesw\immediate\write\csname #1\endcsname{\string\citation{\@citeb}}\fi}%
  \@esphack}
\makeatother 

\begin{document}
\newacronym{3gpp}{3GPP}{3rd Generation Partnership Project}
\newacronym{sriov}{SR-IOV}{Single Root I/O Virtualization}
\newacronym{vf}{VF}{Vitual Functions}
\newacronym{4g}{4G}{4th generation}
\newacronym{5g}{5G}{5th generation}
\newacronym{6g}{6G}{6th generation}
\newacronym{5gc}{5GC}{5G Core}
\newacronym{adc}{ADC}{Analog to Digital Converter}
\newacronym{aerpaw}{AERPAW}{Aerial Experimentation and Research Platform for Advanced Wireless}
\newacronym{ai}{AI}{Artificial Intelligence}
\newacronym{aimd}{AIMD}{Additive Increase Multiplicative Decrease}
\newacronym{am}{AM}{Acknowledged Mode}
\newacronym{amc}{AMC}{Adaptive Modulation and Coding}
\newacronym{amf}{AMF}{Access and Mobility Management Function}
\newacronym{aops}{AOPS}{Adaptive Order Prediction Scheduling}
\newacronym{api}{API}{Application Programming Interface}
\newacronym{apn}{APN}{Access Point Name}
\newacronym{ap}{AP}{Application Protocol}
\newacronym{aqm}{AQM}{Active Queue Management}
\newacronym{ausf}{AUSF}{Authentication Server Function}
\newacronym{avc}{AVC}{Advanced Video Coding}
\newacronym{awgn}{AGWN}{Additive White Gaussian Noise}
\newacronym{balia}{BALIA}{Balanced Link Adaptation Algorithm}
\newacronym{bbu}{BBU}{Base Band Unit}
\newacronym{bdp}{BDP}{Bandwidth-Delay Product}
\newacronym{ber}{BER}{Bit Error Rate}
\newacronym{bf}{BF}{Beamforming}
\newacronym{bler}{BLER}{Block Error Rate}
\newacronym{brr}{BRR}{Bayesian Ridge Regressor}
\newacronym{bs}{BS}{Base Station}
\newacronym{bsr}{BSR}{Buffer Status Report}
\newacronym{bss}{BSS}{Business Support System}
\newacronym{ca}{CA}{Carrier Aggregation}
\newacronym{caas}{CaaS}{Connectivity-as-a-Service}
\newacronym{cb}{CB}{Code Block}
\newacronym{cc}{CC}{Congestion Control}
\newacronym{ccid}{CCID}{Congestion Control ID}
\newacronym{cco}{CC}{Carrier Component}
\newacronym{cd}{CD}{Continuous Deployment}
\newacronym{cdd}{CDD}{Cyclic Delay Diversity}
\newacronym{cdf}{CDF}{Cumulative Distribution Function}
\newacronym{cdn}{CDN}{Content Distribution Network}
\newacronym{cli}{CLI}{Command-line Interface}
\newacronym{cn}{CN}{Core Network}
\newacronym{codel}{CoDel}{Controlled Delay Management}
\newacronym{comac}{COMAC}{Converged Multgi-Access and Core}
\newacronym{cord}{CORD}{Central Office Re-architected as a Datacenter}
\newacronym{cornet}{CORNET}{COgnitive Radio NETwork}
\newacronym{cosmos}{COSMOS}{Cloud Enhanced Open Software Defined Mobile Wireless Testbed for City-Scale Deployment}
\newacronym{cots}{COTS}{Commercial Off-the-Shelf}
\newacronym{cp}{CP}{Control Plane}
\newacronym{cyp}{CP}{Cyclic Prefix}
\newacronym{up}{UP}{User Plane}
\newacronym{cpu}{CPU}{Central Processing Unit}
\newacronym{cqi}{CQI}{Channel Quality Information}
\newacronym{cr}{CR}{Cognitive Radio}
\newacronym{cran}{CRAN}{Cloud RAN}
\newacronym{crs}{CRS}{Cell Reference Signal}
\newacronym{csi}{CSI}{Channel State Information}
\newacronym{csirs}{CSI-RS}{Channel State Information - Reference Signal}
\newacronym{cu}{CU}{Central Unit}
\newacronym{cubb}{cuBB}{CUDA Baseband}
\newacronym{d2tcp}{D$^2$TCP}{Deadline-aware Data center TCP}
\newacronym{d3}{D$^3$}{Deadline-Driven Delivery}
\newacronym{dac}{DAC}{Digital to Analog Converter}
\newacronym{dag}{DAG}{Directed Acyclic Graph}
\newacronym{das}{DAS}{Distributed Antenna System}
\newacronym{dash}{DASH}{Dynamic Adaptive Streaming over HTTP}
\newacronym{dc}{DC}{Dual Connectivity}
\newacronym{dccp}{DCCP}{Datagram Congestion Control Protocol}
\newacronym{dce}{DCE}{Direct Code Execution}
\newacronym{dci}{DCI}{Downlink Control Information}
\newacronym{dctcp}{DCTCP}{Data Center TCP}
\newacronym{dl}{DL}{Downlink}
\newacronym{dmr}{DMR}{Deadline Miss Ratio}
\newacronym{dmrs}{DMRS}{DeModulation Reference Signal}
\newacronym{drlcc}{DRL-CC}{Deep Reinforcement Learning Congestion Control}
\newacronym{drs}{DRS}{Discovery Reference Signal}
\newacronym{du}{DU}{Distributed Unit}
\newacronym{e2e}{E2E}{end-to-end}
\newacronym{earfcn}{EARFCN}{E-UTRA Absolute Radio Frequency Channel Number}
\newacronym{ecaas}{ECaaS}{Edge-Cloud-as-a-Service}
\newacronym{ecn}{ECN}{Explicit Congestion Notification}
\newacronym{edf}{EDF}{Earliest Deadline First}
\newacronym{embb}{eMBB}{Enhanced Mobile Broadband}
\newacronym{empower}{EMPOWER}{EMpowering transatlantic PlatfOrms for advanced WirEless Research}
\newacronym{enb}{eNB}{evolved Node Base}
\newacronym{endc}{EN-DC}{E-UTRAN-\gls{nr} \gls{dc}}
\newacronym{epc}{EPC}{Evolved Packet Core}
\newacronym{eps}{EPS}{Evolved Packet System}
\newacronym{es}{ES}{Edge Server}
\newacronym{etsi}{ETSI}{European Telecommunications Standards Institute}
\newacronym[firstplural=Estimated Times of Arrival (ETAs)]{eta}{ETA}{Estimated Time of Arrival}
\newacronym{eutran}{E-UTRAN}{Evolved Universal Terrestrial Access Network}
\newacronym{faas}{FaaS}{Function-as-a-Service}
\newacronym{fapi}{FAPI}{Functional Application Platform Interface}
\newacronym{fdd}{FDD}{Frequency Division Duplexing}
\newacronym{fdm}{FDM}{Frequency Division Multiplexing}
\newacronym{fdma}{FDMA}{Frequency Division Multiple Access}
\newacronym{fed4fire}{FED4FIRE+}{Federation 4 Future Internet Research and Experimentation Plus}
\newacronym{fir}{FIR}{Finite Impulse Response}
\newacronym{fit}{FIT}{Future \acrlong{iot}}
\newacronym{fpga}{FPGA}{Field Programmable Gate Array}
\newacronym{fr2}{FR2}{Frequency Range 2}
\newacronym{fs}{FS}{Fast Switching}
\newacronym{fscc}{FSCC}{Flow Sharing Congestion Control}
\newacronym{ftp}{FTP}{File Transfer Protocol}
\newacronym{fw}{FW}{Flow Window}
\newacronym{ge}{GE}{Gaussian Elimination}
\newacronym{gnb}{gNB}{Next Generation Node Base}
\newacronym{gop}{GOP}{Group of Pictures}
\newacronym{gpr}{GPR}{Gaussian Process Regressor}
\newacronym{gpu}{GPU}{Graphics Processing Unit}
\newacronym{gtp}{GTP}{GPRS Tunneling Protocol}
\newacronym{gtpc}{GTP-C}{GPRS Tunnelling Protocol Control Plane}
\newacronym{gtpu}{GTP-U}{GPRS Tunnelling Protocol User Plane}
\newacronym{gtpv2c}{GTPv2-C}{\gls{gtp} v2 - Control}
\newacronym{gw}{GW}{Gateway}
\newacronym{harq}{HARQ}{Hybrid Automatic Repeat reQuest}
\newacronym{hetnet}{HetNet}{Heterogeneous Network}
\newacronym{hh}{HH}{Hard Handover}
\newacronym{hol}{HOL}{Head-of-Line}
\newacronym{hqf}{HQF}{Highest-quality-first}
\newacronym{hss}{HSS}{Home Subscription Server}
\newacronym{http}{HTTP}{HyperText Transfer Protocol}
\newacronym{ia}{IA}{Initial Access}
\newacronym{iab}{IAB}{Integrated Access and Backhaul}
\newacronym{ic}{IC}{Incident Command}
\newacronym{ietf}{IETF}{Internet Engineering Task Force}
\newacronym{imsi}{IMSI}{International Mobile Subscriber Identity}
\newacronym{imt}{IMT}{International Mobile Telecommunication}
\newacronym{iot}{IoT}{Internet of Things}
\newacronym{ip}{IP}{Internet Protocol}
\newacronym{itu}{ITU}{International Telecommunication Union}
\newacronym{kepler}{KEPLER}{Kubernetes-based Efficient Power Level Exporter}
\newacronym{kpi}{KPI}{Key Performance Indicator}
\newacronym{kpm}{KPM}{Key Performance Measurement}
\newacronym{kvm}{KVM}{Kernel-based Virtual Machine}
\newacronym{los}{LOS}{Line-of-Sight}
\newacronym{lsm}{LSM}{Link-to-System Mapping}
\newacronym{lstm}{LSTM}{Long Short Term Memory}
\newacronym{lte}{LTE}{Long Term Evolution}
\newacronym{lxc}{LXC}{Linux Container}
\newacronym{m2m}{M2M}{Machine to Machine}
\newacronym{mac}{MAC}{Medium Access Control}
\newacronym{manet}{MANET}{Mobile Ad Hoc Network}
\newacronym{mano}{MANO}{Management and Orchestration}
\newacronym{mc}{MC}{Multi-Connectivity}
\newacronym{mcc}{MCC}{Mobile Cloud Computing}
\newacronym{mchem}{MCHEM}{Massive Channel Emulator}
\newacronym{mcs}{MCS}{Modulation and Coding Scheme}
\newacronym{mec}{MEC}{Multi-access Edge Computing}
\newacronym{mec2}{MEC}{Mobile Edge Cloud}
\newacronym{mfc}{MFC}{Mobile Fog Computing}
\newacronym{mgen}{MGEN}{Multi-Generator}
\newacronym{mi}{MI}{Mutual Information}
\newacronym{mib}{MIB}{Master Information Block}
\newacronym{miesm}{MIESM}{Mutual Information Based Effective SINR}
\newacronym{mimo}{MIMO}{Multiple Input, Multiple Output}
\newacronym{ml}{ML}{Machine Learning}
\newacronym{mlr}{MLR}{Maximum-local-rate}
\newacronym[plural=\gls{mme}s,firstplural=Mobility Management Entities (MMEs)]{mme}{MME}{Mobility Management Entity}
\newacronym{mmtc}{mMTC}{Massive Machine-Type Communications}
\newacronym{mmwave}{mmWave}{millimeter wave}
\newacronym{mpdccp}{MP-DCCP}{Multipath Datagram Congestion Control Protocol}
\newacronym{mptcp}{MPTCP}{Multipath TCP}
\newacronym{mr}{MR}{Maximum Rate}
\newacronym{mrdc}{MR-DC}{Multi \gls{rat} \gls{dc}}
\newacronym{mse}{MSE}{Mean Square Error}
\newacronym{mss}{MSS}{Maximum Segment Size}
\newacronym{mt}{MT}{Mobile Termination}
\newacronym{mtd}{MTD}{Machine-Type Device}
\newacronym{mtu}{MTU}{Maximum Transmission Unit}
\newacronym{mumimo}{MU-MIMO}{Multi-user \gls{mimo}}
\newacronym{mvno}{MVNO}{Mobile Virtual Network Operator}
\newacronym{nalu}{NALU}{Network Abstraction Layer Unit}
\newacronym{nas}{NAS}{Network Attached Storage}
\newacronym{nat}{NAT}{Network Address Translation}
\newacronym{nbiot}{NB-IoT}{Narrow Band IoT}
\newacronym{nfv}{NFV}{Network Function Virtualization}
\newacronym{nfvi}{NFVI}{Network Function Virtualization Infrastructure}
\newacronym{ni}{NI}{Network Interfaces}
\newacronym{llm}{LLM}{Large Language Model}
\newacronym{dpdk}{DPDK}{Data Plane Development Kit}
\newacronym{cicd}{CI/CD}{Continuous Integration and Continuous Delivery/Deployment}
\newacronym{nic}{NIC}{Network Interface Card}
\newacronym{nlos}{NLOS}{Non-Line-of-Sight}
\newacronym{now}{NOW}{Non Overlapping Window}
\newacronym{nsm}{NSM}{Network Service Mesh}
\newacronym[type=hidden]{nr}{NR}{New Radio}
\newacronym[type=hidden]{ota}{OTA}{Over-The-Air}
\newacronym{nrf}{NRF}{Network Repository Function}
\newacronym{nsa}{NSA}{Non Stand Alone}
\newacronym{nse}{NSE}{Network Slicing Engine}
\newacronym{nssf}{NSSF}{Network Slice Selection Function}
\newacronym{o2i}{O2I}{Outdoor to Indoor}
\newacronym{oai}{OAI}{OpenAirInterface}
\newacronym{oaicn}{OAI-CN}{\gls{oai} \acrlong{cn}}
\newacronym{oairan}{OAI-RAN}{\acrlong{oai} \acrlong{ran}}
\newacronym{oam}{OAM}{Operations, Administration and Maintenance}
\newacronym{ofdm}{OFDM}{Orthogonal Frequency Division Multiplexing}
\newacronym{olia}{OLIA}{Opportunistic Linked Increase Algorithm}
\newacronym{omec}{OMEC}{Open Mobile Evolved Core}
\newacronym{onap}{ONAP}{Open Network Automation Platform}
\newacronym{onf}{ONF}{Open Networking Foundation}
\newacronym{onos}{ONOS}{Open Networking Operating System}
\newacronym{oom}{OOM}{\gls{onap} Operations Manager}
\newacronym{opnfv}{OPNFV}{Open Platform for \gls{nfv}}
\newacronym{orbit}{ORBIT}{Open-Access Research Testbed for Next-Generation Wireless Networks}
\newacronym{os}{OS}{Operating System}
\newacronym{oss}{OSS}{Operations Support System}
\newacronym{pa}{PA}{Position-aware}
\newacronym{pase}{PASE}{Prioritization, Arbitration, and Self-adjusting Endpoints}
\newacronym{pawr}{PAWR}{Platforms for Advanced Wireless Research}
\newacronym{pbch}{PBCH}{Physical Broadcast Channel}
\newacronym{pcef}{PCEF}{Policy and Charging Enforcement Function}
\newacronym{pcfich}{PCFICH}{Physical Control Format Indicator Channel}
\newacronym{pcrf}{PCRF}{Policy and Charging Rules Function}
\newacronym{pdcch}{PDCCH}{Physical Downlink Control Channel}
\newacronym{pdcp}{PDCP}{Packet Data Convergence Protocol}
\newacronym{pdf}{PDF}{Probability Density Function}
\newacronym{pdsch}{PDSCH}{Physical Downlink Shared Channel}
\newacronym{pdu}{PDU}{Packet Data Unit}
\newacronym{pf}{PF}{Proportional Fair}
\newacronym{pgw}{PGW}{Packet Gateway}
\newacronym{phich}{PHICH}{Physical Hybrid ARQ Indicator Channel}
\newacronym{phy}{PHY}{Physical}
\newacronym{pmch}{PMCH}{Physical Multicast Channel}
\newacronym{pmi}{PMI}{Precoding Matrix Indicators}
\newacronym{powder}{POWDER}{Platform for Open Wireless Data-driven Experimental Research}
\newacronym{ppo}{PPO}{Proximal Policy Optimization}
\newacronym{ppp}{PPP}{Poisson Point Process}
\newacronym{prach}{PRACH}{Physical Random Access Channel}
\newacronym{prb}{PRB}{Physical Resource Block}
\newacronym{psnr}{PSNR}{Peak Signal to Noise Ratio}
\newacronym{pss}{PSS}{Primary Synchronization Signal}
\newacronym{pucch}{PUCCH}{Physical Uplink Control Channel}
\newacronym{pusch}{PUSCH}{Physical Uplink Shared Channel}
\newacronym{qam}{QAM}{Quadrature Amplitude Modulation}
\newacronym{qci}{QCI}{\gls{qos} Class Identifier}
\newacronym{qoe}{QoE}{Quality of Experience}
\newacronym{qos}{QoS}{Quality of Service}
\newacronym{quic}{QUIC}{Quick UDP Internet Connections}
\newacronym{rach}{RACH}{Random Access Channel}
\newacronym{ran}{RAN}{Radio Access Network}
\newacronym[firstplural=Radio Access Technologies (RATs)]{rat}{RAT}{Radio Access Technology}
\newacronym{rbg}{RBG}{Resource Block Group}
\newacronym{rcn}{RCN}{Research Coordination Network}
\newacronym{rc}{RC}{RAN Control}
\newacronym{rec}{REC}{Radio Edge Cloud}
\newacronym{red}{RED}{Random Early Detection}
\newacronym{renew}{RENEW}{Reconfigurable Eco-system for Next-generation End-to-end Wireless}
\newacronym{rf}{RF}{Radio Frequency}
\newacronym{rfc}{RFC}{Request for Comments}
\newacronym{rfr}{RFR}{Random Forest Regressor}
\newacronym{ric}{RIC}{RAN Intelligent Controller}
\newacronym{nrric}{Near-RT RIC}{Near-Real-Time RAN Intelligent Controller}
\newacronym{rlc}{RLC}{Radio Link Control}
\newacronym{rlf}{RLF}{Radio Link Failure}
\newacronym{rlnc}{RLNC}{Random Linear Network Coding}
\newacronym{rmr}{RMR}{RIC Message Router}
\newacronym{rmse}{RMSE}{Root Mean Squared Error}
\newacronym{rnis}{RNIS}{Radio Network Information Service}
\newacronym{rr}{RR}{Round Robin}
\newacronym{rrc}{RRC}{Radio Resource Control}
\newacronym{rrm}{RRM}{Radio Resource Management}
\newacronym{rru}{RRU}{Remote Radio Unit}
\newacronym{rs}{RS}{Remote Server}
\newacronym{rsrp}{RSRP}{Reference Signal Received Power}
\newacronym{rsrq}{RSRQ}{Reference Signal Received Quality}
\newacronym{rss}{RSS}{Received Signal Strength}
\newacronym{rssi}{RSSI}{Received Signal Strength Indicator}
\newacronym{rtt}{RTT}{Round Trip Time}
\newacronym{ru}{RU}{Radio Unit}
\newacronym{rw}{RW}{Receive Window}
\newacronym{rx}{RX}{Receiver}
\newacronym{s1ap}{S1AP}{S1 Application Protocol}
\newacronym{sa}{SA}{standalone}
\newacronym{sack}{SACK}{Selective Acknowledgment}
\newacronym{sap}{SAP}{Service Access Point}
\newacronym{sc2}{SC2}{Spectrum Collaboration Challenge}
\newacronym{scef}{SCEF}{Service Capability Exposure Function}
\newacronym{sch}{SCH}{Secondary Cell Handover}
\newacronym{scoot}{SCOOT}{Split Cycle Offset Optimization Technique}
\newacronym{sctp}{SCTP}{Stream Control Transmission Protocol}
\newacronym{sdap}{SDAP}{Service Data Adaptation Protocol}
\newacronym{sdk}{SDK}{Software Development Kit}
\newacronym{sdm}{SDM}{Space Division Multiplexing}
\newacronym{sdma}{SDMA}{Spatial Division Multiple Access}
\newacronym{sdl}{SDL}{Shared Data Layer}
\newacronym{sdn}{SDN}{Software-defined Networking}
\newacronym{sdr}{SDR}{Software-defined Radio}
\newacronym{seba}{SEBA}{SDN-Enabled Broadband Access}
\newacronym{sgsn}{SGSN}{Serving GPRS Support Node}
\newacronym{sgw}{SGW}{Service Gateway}
\newacronym{si}{SI}{Study Item}
\newacronym{sib}{SIB}{Secondary Information Block}
\newacronym{sinr}{SINR}{Signal to Interference plus Noise Ratio}
\newacronym{sip}{SIP}{Session Initiation Protocol}
\newacronym{siso}{SISO}{Single Input, Single Output}
\newacronym{sla}{SLA}{Service Level Agreement}
\newacronym{sm}{SM}{Service Model}
\newacronym{e2sm}{E2SM}{E2 Service Model}
\newacronym{e2ap}{E2AP}{E2 Application Protocol}
\newacronym{smf}{SMF}{Session Management Function}
\newacronym{smo}{SMO}{Service Management and Orchestration}
\newacronym{sms}{SMS}{Short Message Service}
\newacronym{smsgmsc}{SMS-GMSC}{\gls{sms}-Gateway}
\newacronym{snr}{SNR}{Signal-to-Noise-Ratio}
\newacronym{son}{SON}{Self-Organizing Network}
\newacronym{sptcp}{SPTCP}{Single Path TCP}
\newacronym{srb}{SRB}{Service Radio Bearer}
\newacronym{srn}{SRN}{Standard Radio Node}
\newacronym{srs}{SRS}{Sounding Reference Signal}
\newacronym{ss}{SS}{Synchronization Signal}
\newacronym{sss}{SSS}{Secondary Synchronization Signal}
\newacronym{st}{ST}{Spanning Tree}
\newacronym{svc}{SVC}{Scalable Video Coding}
\newacronym{tb}{TB}{Transport Block}
\newacronym{tcp}{TCP}{Transmission Control Protocol}
\newacronym{tdd}{TDD}{Time Division Duplexing}
\newacronym{tdm}{TDM}{Time Division Multiplexing}
\newacronym{tdma}{TDMA}{Time Division Multiple Access}
\newacronym{tfl}{TfL}{Transport for London}
\newacronym{tfrc}{TFRC}{TCP-Friendly Rate Control}
\newacronym{tft}{TFT}{Traffic Flow Template}
\newacronym{tgen}{TGEN}{Traffic Generator}
\newacronym{tip}{TIP}{Telecom Infra Project}
\newacronym{tm}{TM}{Transparent Mode}
\newacronym{to}{TO}{Telco Operator}
\newacronym{tr}{TR}{Technical Report}
\newacronym{trp}{TRP}{Transmitter Receiver Pair}
\newacronym{ts}{TS}{Technical Specification}
\newacronym{tti}{TTI}{Transmission Time Interval}
\newacronym{ttt}{TTT}{Time-to-Trigger}
\newacronym{tx}{TX}{Transmitter}
\newacronym{uas}{UAS}{Unmanned Aerial System}
\newacronym{uav}{UAV}{Unmanned Aerial Vehicle}
\newacronym{udm}{UDM}{Unified Data Management}
\newacronym{udp}{UDP}{User Datagram Protocol}
\newacronym{udr}{UDR}{Unified Data Repository}
\newacronym{ue}{UE}{User Equipment}
\newacronym{uhd}{UHD}{\gls{usrp} Hardware Driver}
\newacronym{ul}{UL}{Uplink}
\newacronym{um}{UM}{Unacknowledged Mode}
\newacronym{uml}{UML}{Unified Modeling Language}
\newacronym{upa}{UPA}{Uniform Planar Array}
\newacronym{upf}{UPF}{User Plane Function}
\newacronym{urllc}{URLLC}{Ultra Reliable and Low Latency Communications}
\newacronym{usa}{U.S.}{United States}
\newacronym{usim}{USIM}{Universal Subscriber Identity Module}
\newacronym{usrp}{USRP}{Universal Software Radio Peripheral}
\newacronym{utc}{UTC}{Urban Traffic Control}
\newacronym{vim}{VIM}{Virtualization Infrastructure Manager}
\newacronym{vm}{VM}{Virtual Machine}
\newacronym{vnf}{VNF}{Virtual Network Function}
\newacronym{volte}{VoLTE}{Voice over \gls{lte}}
\newacronym{voltha}{VOLTHA}{Virtual OLT HArdware Abstraction}
\newacronym{vr}{VR}{Virtual Reality}
\newacronym{vran}{vRAN}{Virtualized RAN}
\newacronym{vss}{VSS}{Video Streaming Server}
\newacronym{wbf}{WBF}{Wired Bias Function}
\newacronym{wf}{WF}{Waterfilling}
\newacronym{wg}{WG}{Working Group}
\newacronym{wlan}{WLAN}{Wireless Local Area Network}
\newacronym{osm}{OSM}{Open Source \gls{nfv} Management and Orchestration}
\newacronym{pnf}{PNF}{Physical Network Function}
\newacronym{drl}{DRL}{Deep Reinforcement Learning}
\newacronym{mtc}{MTC}{Machine-type Communications}
\newacronym{osc}{OSC}{O-RAN Software Community}
\newacronym{mns}{MnS}{Management Services}
\newacronym{ves}{VES}{\gls{vnf} Event Stream}
\newacronym{ei}{EI}{Enrichment Information}
\newacronym{fh}{FH}{Fronthaul}
\newacronym{fft}{FFT}{Fast Fourier Transform}
\newacronym{laa}{LAA}{Licensed-Assisted Access}
\newacronym{plfs}{PLFS}{Physical Layer Frequency Signals}
\newacronym{ptp}{PTP}{Precision Time Protocol}
\newacronym{ntp}{NTP}{Network Time Protocol}
\newacronym{cbrs}{CBRS}{Citizen Broadband Radio Service}
\newacronym{rnti}{RNTI}{Radio Network Temporary Identifier}
\newacronym{tbs}{TBS}{Transport Block Size}
\newacronym{nfd}{NFD}{Node Feature Discovery}
\newacronym{mcp}{MCP}{Machine Configuration Pool}
\newacronym{vpn}{VPN}{Virtual Private Network}
\newacronym{onr}{ONR}{Office of Naval Research}
\newacronym{afosr}{AFOSR}{Air Force Office of Scientific Research}
\newacronym{afrl}{AFRL}{Air Force Research Laboratory}
\newacronym{arl}{ARL}{Army Research Laboratory}

\newacronym{arc}{ARC}{Aerial Research Cloud}

\newacronym{ct}{CT}{Continuous Testing}
\newacronym{mno}{MNO}{Mobile Network Operator}
\newacronym{oci}{OCI}{Open Container Initiative}
\newacronym{macsec}{MACsec}{Media Access Control Security}
\newacronym{pt}{PT}{Plain Text}
\newacronym{cuda}{CUDA}{Compute Unified Device Architecture}
\newacronym{dsp}{DSP}{Digital Signal Processing}

\newacronym{cus}{CUS}{Control, User, Synchronization}
\newacronym{dpd}{DPD}{Digital Pre-Distorsion}
\newacronym{cfr}{CFR}{Crest Factor Reduction}
\newacronym{pci}{PCIe}{Peripheral Component Interconnect Express}
\newacronym{dpu}{DPU}{Data Processing Unit}
\newacronym{rfsoc}{RFSoC}{Radio Frequency System-on-Chip}
\newacronym{if}{IF}{Intermediate Frequency}
\newacronym{nyu}{NYU}{New York University}
\newacronym{gh}{GH}{Grace Hopper}
\newacronym{trl}{TRL}{Technology Readiness Level}
\newacronym{srfa}{SRFA}{Special Research Focus Area}
\newacronym{qsfp}{QSFP}{quad small form factor pluggable}
\newacronym{pse}{PSE}{Performance Specialized Engine}
\newacronym{cae}{CAE}{Cognitive Analysis Engine}
\newacronym{simd}{SIMD}{Single Instruction/Multiple Data}
\newacronym{rt}{RT}{Real-Time}
\newacronym{asm}{ASM}{Advanced Sleep Mode}
\newacronym{aoa}{AoA}{Angle of Arrival}
\newacronym{eaxcid}{eAxC\_ID}{extended Antenna-Carrier Identifier}
\newacronym{bwp}{BWP}{Bandwidth Part}
\newacronym{dfe}{DFE}{Digital Front-End}
\newacronym{spi}{SPI}{Serial Peripheral Interface}
\newacronym{gpio}{GPIO}{General Purpose Input/Output}
\newacronym{nco}{NCO}{Numerically Controlled Oscillator}
\newacronym{lo}{LO}{Local Oscillator}
\newacronym{lna}{LNA}{Low-Noise Amplifier}
\newacronym{pll}{PLL}{Phased-Locked Loop}
\newacronym{som}{SOM}{System-on-Module}
\newacronym{papr}{PAPR}{Peak-to-Average Power Ratio}
\newacronym{pcb}{PCB}{Printed Circuit Board}
\newacronym{gcpw}{GCPW}{Grounded Co-Planar Waveguide}
\newacronym{cnn}{CNN}{Convolutional Neural Network}
\newacronym{gmp}{GMP}{Generalized Memory Polynomial}
\newacronym{ngrg}{nGRG}{next Generation Research Group}
\newacronym{mrl}{MRL}{Manufacturing Readiness Level}
\newacronym{fr}{FR}{Frequency Range}

\newacronym{sbom}{SBOM}{Software Bill of Materials}
\newacronym{hbom}{HBOM}{Hardware Bill of Materials}
\newacronym{vex}{VEX}{Vulnerability Exploitability eXchange}
\newacronym{dos}{DoS}{Denial of Service}
\newacronym{sme}{SME}{Small-Medium Enterprise}

\newacronym{ulpi}{ULPI}{Uplink Performance Improvement}
\newacronym{oem}{OEM}{Original Equipment Manufacturer}
\newacronym{nsin}{NSIN}{National Security Innovation Network}
\newacronym{dod}{DoD}{Department of Defense}
\newacronym{arpu}{ARPU}{Average Revenue per User}
\newacronym{opex}{OPEX}{operational expenses}
\newacronym{txb}{TXB}{Transmit Beam}
\newacronym{cve}{CVE}{Common Vulnerabilities and Exposure}
\newacronym{json}{JSON}{JavaScript Object Notation}
\newacronym{it}{IT}{Information Technology}
\newacronym{ci}{CI}{Continuous Integration}
\newacronym{nlp}{NLP}{Natural Language Processing}
\newacronym{sba}{SBA}{Service Based Architecture}
\newacronym{raid}{RAID}{Redundant Array of Independent Disks}
\newacronym{naturallanguage}{NL}{Natural Language}
\newacronym{lan}{LAN}{Local Area Network}
\newacronym{ipc}{IPC}{Inter-process Communication}
\newacronym{adb}{ADB}{Android Debug Bridge}
\newacronym{mig}{MIG}{Multi-Instance GPU}
\tikzstyle{startstop} = [rectangle, rounded corners, minimum width=2cm, minimum height=0.5cm,text centered, draw=black]
\tikzstyle{io} = [trapezium, trapezium left angle=70, trapezium right angle=110, minimum width=3cm, minimum height=1cm, text centered, draw=black]
\tikzstyle{process} = [rectangle, minimum width=2cm, minimum height=0.5cm, text centered, draw=black, alignb=center]
\tikzstyle{decision} = [ellipse, minimum width=2cm, minimum height=1cm, text centered, draw=black]
\tikzstyle{arrow} = [thick,<->,>=stealth]
\tikzstyle{line} = [thick,>=stealth]
\tikzstyle{darrow} = [thick,<->,>=stealth,dashed]
\tikzstyle{sarrow} = [thick,->,>=stealth]
\tikzstyle{larrow} = [line width=0.1mm,dashdotted,->,>=stealth]
\tikzstyle{llarrow} = [line width=0.1mm,->,>=stealth]

\makeatletter
\def\grd@save@target#1{%
  \def\grd@target{#1}}
\def\grd@save@start#1{%
  \def\grd@start{#1}}
\tikzset{
  grid with coordinates/.style={
    to path={%
      \pgfextra{%
        \edef\grd@@target{(\tikztotarget)}%
        \tikz@scan@one@point\grd@save@target\grd@@target\relax
        \edef\grd@@start{(\tikztostart)}%
        \tikz@scan@one@point\grd@save@start\grd@@start\relax
        \draw[minor help lines] (\tikztostart) grid (\tikztotarget);
        \draw[major help lines] (\tikztostart) grid (\tikztotarget);
        \grd@start
        \pgfmathsetmacro{\grd@xa}{\the\pgf@x/1cm}
        \pgfmathsetmacro{\grd@ya}{\the\pgf@y/1cm}
        \grd@target
        \pgfmathsetmacro{\grd@xb}{\the\pgf@x/1cm}
        \pgfmathsetmacro{\grd@yb}{\the\pgf@y/1cm}
        \pgfmathsetmacro{\grd@xc}{\grd@xa + \pgfkeysvalueof{/tikz/grid with coordinates/major step x}}
        \pgfmathsetmacro{\grd@yc}{\grd@ya + \pgfkeysvalueof{/tikz/grid with coordinates/major step y}}
        \foreach \x in {\grd@xa,\grd@xc,...,\grd@xb}
        \node[anchor=north] at (\x,\grd@ya) {\pgfmathprintnumber{\x}};
        \foreach \y in {\grd@ya,\grd@yc,...,\grd@yb}
        \node[anchor=east] at (\grd@xa,\y) {\pgfmathprintnumber{\y}};
      }
    }
  },
  minor help lines/.style={
    help lines,
    gray,
    line cap =round,
    xstep=\pgfkeysvalueof{/tikz/grid with coordinates/minor step x},
    ystep=\pgfkeysvalueof{/tikz/grid with coordinates/minor step y}
  },
  major help lines/.style={
    help lines,
    line cap =round,
    line width=\pgfkeysvalueof{/tikz/grid with coordinates/major line width},
    xstep=\pgfkeysvalueof{/tikz/grid with coordinates/major step x},
    ystep=\pgfkeysvalueof{/tikz/grid with coordinates/major step y}
  },
  grid with coordinates/.cd,
  minor step x/.initial=.5,
  minor step y/.initial=.2,
  major step x/.initial=1,
  major step y/.initial=1,
  major line width/.initial=1pt,
}
\makeatother

\title{AutoRAN: Automated and\\Zero-Touch Open RAN Systems}

\author{\IEEEauthorblockN{
		Stefano Maxenti,~\IEEEmembership{Student Member, IEEE},
		Ravis Shirkhani,~\IEEEmembership{Student Member, IEEE},\\
		Maxime Elkael,~\IEEEmembership{Member, IEEE},
		Leonardo Bonati,~\IEEEmembership{Member, IEEE},
		Salvatore D'Oro,~\IEEEmembership{Member, IEEE},\\
		Tommaso Melodia,~\IEEEmembership{Fellow, IEEE},
		Michele Polese,~\IEEEmembership{Senior Member, IEEE}}
	\thanks{S. Maxenti, R. Shirkhani, M. Elkael, L. Bonati, T. Melodia, and M. Polese are with the Institute for the Wireless Internet of Things, Northeastern University, Boston, MA, U.S.A. E-mail: \{maxenti.s, shirkhani.r, m.elkael, l.bonati, melodia, m.polese\}@northeastern.edu. Salvatore D'Oro is with zTouch Networks, Inc., Boston, MA, U.S.A. Email: salvo@ztouchnet.com.}
	\thanks{This work was partially supported by the National Telecommunications and Information Administration (NTIA)'s Public Wireless Supply Chain Innovation Fund (PWSCIF) under Award No. 25-60-IF054 and by the U.S. National Science Foundation under grant CNS-2117814.}
}

\maketitle


\begin{abstract}
\label{abstract}%
Modern cellular networks adopt a software-based and disaggregated approach to support diverse requirements and mission-critical reliability needs.  
While softwarization introduces flexibility, it also increases the complexity of the network architectures, which calls for robust automation frameworks that can deliver efficient and fully-autonomous configuration, scalability, and multi-vendor integration. This paper presents \framework, an automated, intent-driven framework for zero-touch provisioning of open, programmable cellular networks. Leveraging cloud-native principles, \framework employs virtualization, declarative infrastructure-as-code templates, and disaggregated micro-services to abstract physical resources and protocol stacks. Its orchestration engine integrates \glspl{llm} to translate high-level intents into machine-readable configurations, enabling closed-loop control via telemetry-driven observability. Implemented on a multi-architecture OpenShift cluster with heterogeneous compute (x86/ARM CPUs, NVIDIA GPUs) and multi-vendor \gls{ran} hardware (Foxconn, NI), AutoRAN automates deployment of O-RAN-compliant stacks—including OpenAirInterface, NVIDIA ARC \gls{ran}, Open5GS core, and \gls{osc} RIC components—using \gls{cicd} pipelines. Experimental results demonstrate that \framework is capable of deploying an end-to-end Private 5G network in less than {60} seconds with {1.6}\:Gbps throughput, validating its ability to streamline configuration, accelerate testing, and reduce manual intervention with similar performance than non cloud-based implementations. With its novel LLM-assisted intent translation mechanism, and performance-optimized automation workflow for multi-vendor environments, 
\framework has the potential of advancing the robustness of next-generation cellular supply chains through reproducible, intent-based provisioning across public and private deployments.

\end{abstract}


\glsresetall


\vspace{-15pt}
\section{Introduction}
\label{sec:intro}%

Today's cellular networks serve a variety of customers and use cases with heterogeneous deployments and technologies that operate under diverse user requirements and channel conditions~\cite{narayanan2022comparative}. 
This diversity of requirements and strategic support for society and economy make cellular networks extremely complex systems. 
\revision{For instance,} an end-to-end deployment of a \gls{5g} cellular system counts tens of micro-services for the core network, a distributed \gls{ran} with a disaggregated protocol stack capable of handling hundreds of users from a single base station, and additional services for management and optimization of the network~\cite{dahlman20205g}. 

Even carrier networks, where operators \revision{bear} the know-how to manage and control such systems, \revision{suffer from} outages and anomalies due to the complex nature of cellular networks~\cite{outage-report}.
This major pain point is further exacerbated in private \gls{5g} networks, often operated by enterprise \gls{it} departments, which face challenges in reliably planning, deploying, operating, and scaling private connectivity solutions. Compared to Wi-Fi, which is usually deployed as integrated solution with simple and inexpensive access points, private cellular systems can provide the necessary performance guarantees to enable mission-critical use cases, and deliver the ultra-low latency and high throughput connectivity needed for enabling Industry 4.0 and automation~\cite{industry4.0}. On the other hand, cellular systems require expert knowledge and a more involved management and monitoring effort. For this reason, enterprise often resorts to system integrators, which streamline the deployment and management process but add an intermediate layer between the network and the enterprise. This also increases integration costs \cite{system-integration} and reduces the ability to customize the network.
Part of this complexity also stems from the recent transition of cellular systems to software-based and disaggregated architectures such as \gls{cran}, \gls{vran} and Open \gls{ran}. Indeed, this transition brings flexibility and programmability, as well as support for multi-vendor deployments. However, the decoupling of \gls{ran} elements results in an increased number of network functions (e.g, \gls{ru}, \gls{du} and \gls{cu} in a disaggregated \gls{gnb}) that add complexity to the network and call for proper automation tools capable of taming such complexity. Indeed, while \cicd and automation are widely used in cloud solutions, existing techniques cannot be directly applied to cellular systems due to challenges involving radio transmissions, spectrum allocation policies, distributed deployments, core and \gls{ran} elements for low latency, high throughput, and the support of the real-time processing for wireless signals.
\revision{Notably, generic cloud platforms assume best-effort workloads and lack support for RAN-specific requirements such as CPU pinning, nanosecond clock synchronization, and specialized networking.}

In this paper, we address the above challenges and propose, design, and develop \framework, an automated and intelligent framework for zero-touch configuration and provisioning of open and programmable cellular networks. \revision{\framework extends cloud-native abstractions to provide support for \gls{ran}-specific workloads,} with automation and virtualization that extend from the core network to the cell site.

\begin{figure}[t]
    \centering
    \includegraphics[width=\columnwidth]{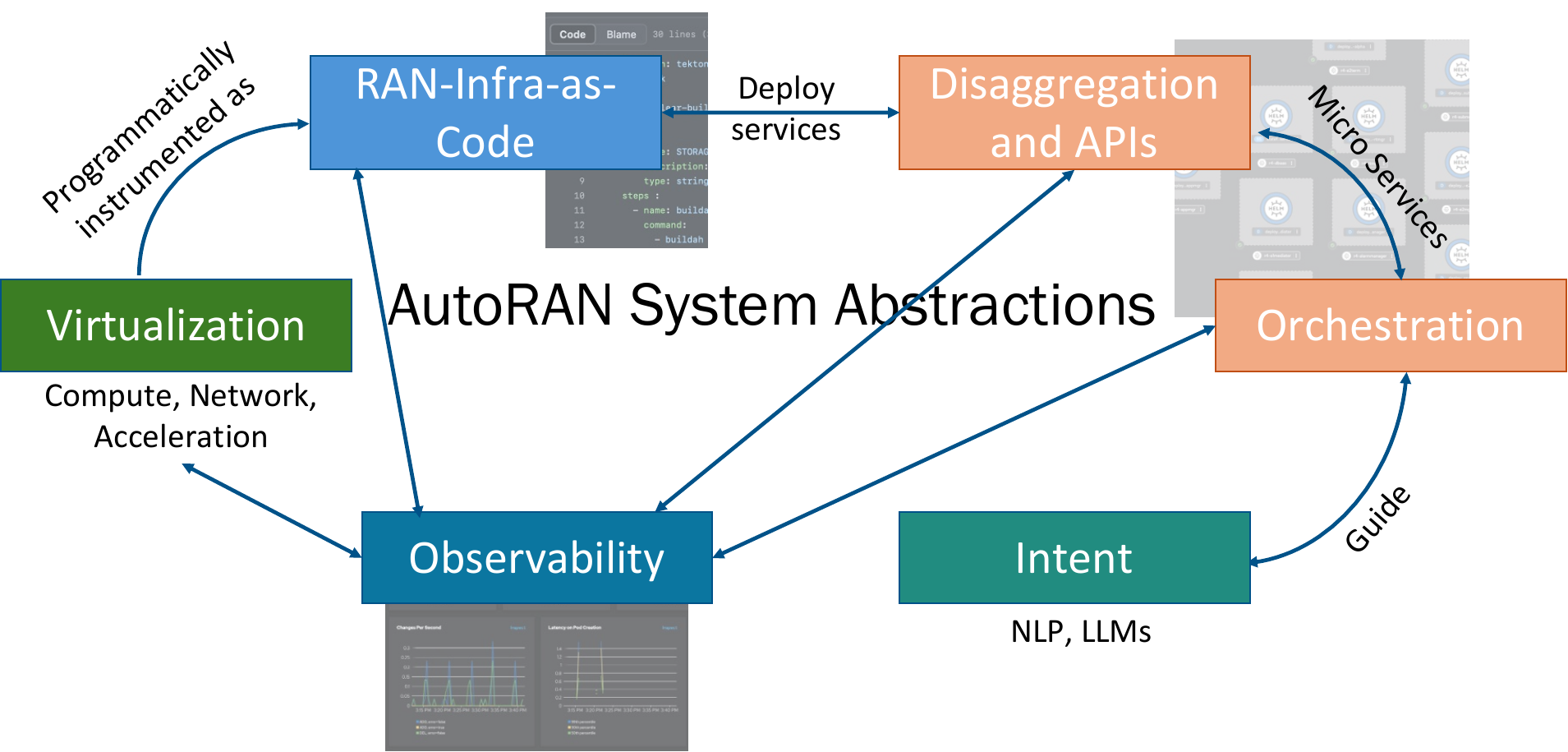}
    \caption{Key abstractions for the \framework system.}
    \label{fig:abstraction}
    \vspace{-.6cm}
\end{figure}
The \framework design (Figure~\ref{fig:abstraction}) is based on key abstractions and capabilities: (i) the physical infrastructure is abstracted through virtualization of networking, hardware accelerators, and compute resources. This is programmatically configured and operationalized using a (ii) declarative approach, where code and templates are used to describe the characteristics and configuration of the system, enabling versioning control, automated configuration roll-outs, and end-to-end optimization of infrastructure parameters. The applications, which in this case are \ran-related workloads, are deployed as (iii) disaggregated micro-services, coexisting on the same infrastructure as traditional IT and cloud deployments, while exposing \glspl{api} for reconfiguration, interaction across services, and telemetry. For example, a base station is split into multiple micro-services (i.e., \gls{du} and \gls{cu}) connected through open interfaces. The service life-cycle is managed through (iv) end-to-end orchestration to automatically converge to a valid set of services with the appropriate status and configuration across software applications and infrastructure. The orchestrator is guided by (v) high-level intents for operator requirements (e.g., coverage area, minimum \gls{qos} level) and translated into machine-readable configurations through \gls{nlp} and \glspl{llm}. Finally, (vi) observability makes it possible to track the end-to-end system status and coordinate with the orchestrator to implement closed-loop control.

We designed and implemented a system that abides by such principles on a fully-programmable OpenShift cluster \cite{openshift} with heterogeneous compute architectures, accelerators, and end-to-end components for the software stacks. Specifically, we design a \rifc solution that, based on an intent expressed by the user through an \gls{llm} prompt, deploys and configures \gls{gnb} on a cluster with 13 nodes with x86 and ARM \glspl{cpu}, NVIDIA L40, A100, and GH200 \glspl{gpu}, and radios from various radio manufacturers (Foxconn and NI/Ettus). 
The system automation combines OpenShift, Tekton pipelines, ArgoCD, and similar tools to automatically deploy an optimized \gls{ran} based on \gls{oai}~\cite{kaltenberger2024driving, nikaein2014openairinterface}, on NVIDIA ARC-OTA~\cite{villa2024x5g, aerial}, or on srsRAN~\cite{srsran}, core network based on Open5GS, and \gls{ric} components from the \gls{osc}. 
Once the system is deployed, the automation framework orchestrates end-to-end performance and functional tests.
We show how to transition from generic bare metal deployments to integrated and automated deployments on clusterization platforms.

Overall, the contributions of our paper are as follows:
\begin{itemize}[leftmargin=*]
    \item Conceptualize, implement and evaluate \framework, an end-to-end automation solution for zero-touch deployment, configuration and testing of multi-vendor cellular networks, taking advantage of hardware accelerators integrated into a cloud computing platform;
    \item We efficiently integrate various \gls{ran} stacks into a unified computing and automation platform, introducing advanced virtualization techniques and showing \gls{ran} coexistence with generic cloud computing software;
    \item Automate deployment and testing with \framework to achieve an end-to-end deployment in less than 60 seconds and peak throughput of $1.6$\:Gbps on the cloud-based \framework infrastructure;
    \item Design and develop an \gls{llm} pipeline that converts high-level intents into verifiable and bespoke cluster configuration and \ran deployment policies; 
    \item Extensively profile \framework performance on different tasks, including deploying and testing cellular networks.
\end{itemize}

The remainder of this paper is organized as follows. 
In Section~\ref{sec:related_works}, we review literature works related to ours.
In Section~\ref{sec:e2e-telco-cloud}, we discuss the foundational design principles behind \framework, while
in Section~\ref{sec:integration_details} we discuss the implementation of such principles on the cluster infrastructure. 
In Section~\ref{sec:automated-workflows}, we focus on the automation workflows for deployment and testing and describe the design and training procedures of our \gls{llm}. 
Finally, in Section~\ref{sec:experimental-evaluation}, we provide results and metrics, while in Section~\ref{sec:conclusions}, we draw our conclusions and discuss future works.
\section{Related Work}%
\label{sec:related_works}
\framework proposes a solution to automatically configure, deploy, and operate a multi-vendor Open \gls{ran} system.
It also provides a convenient way to perform repeatable tests with different specifications, and deployment through a \gls{llm} interface.
The work encompasses various aspects of configuring infrastructure for automated deployment and testing of Open \gls{ran} while investigating \gls{ai} applications in telecommunications.

The authors of~\cite{microsoft-testbed} introduce an enterprise-scale Open \gls{ran} testbed that enables realistic, high-fidelity research. By automating Open \gls{ran} deployment and real-time telemetry collection, they aim at enabling testing and optimization of network functions. 
NeutRAN~\cite{bonati2023neutran} provides zero-touch multitenancy through \gls{ran}/spectrum sharing on OpenShift, dynamically allocating resources via optimization rApps for efficient utilization. Similarly, 5G-CT~\cite{bonati20235gct} automates \gls{e2e} 5G/O-RAN networks using OpenShift and GitOps workflows, integrating capabilities for the continuous integration, deployment, and testing with \gls{oai}, commercial core, and \glspl{sdr}.
5GShell~\cite{5gshell} aims at reducing human-based network configurations by providing a plug-and-play framework designed to automate the deployment of 5G cellular networks. It enables users to deploy different cores, two different protocol stacks, and softwarized \gls{ue}.
A demo of cloud-native 5G network automation using Kubernetes and OpenShift operators is presented in~\cite{5g-cloud-native-demo}. This solution automates deployment, configuration, service upgrade, and switching between monolithic and disaggregated \gls{ran} architectures based on network traffic. 
Although these works focus on automating cloud computing and virtualization for Open \gls{ran}, they do not include hardware accelerators such as NVIDIA GPUs in the infrastructure. \framework, instead, has a more diverse and heterogeneous infrastructure both for \glspl{ru} and \gls{gnb} stacks, and is built with cloud-computing automation and scalability in mind.

\gls{ran} accelerators are instead considered in~\cite{villa2024x5g,lls-cloudric}. X5G~\cite{villa2024x5g} is an open, programmable, and multi-vendor private 5G testbed that integrates NVIDIA GPUs to accelerate the 5G physical layer interfacing with \gls{oai} for high-\gls{du}. CloudRIC~\cite{lls-cloudric}, is a virtualized O-RAN solution that reduces cost and improves energy efficiency by pooling hardware accelerators such as NVIDIA GPUs and FPGAs across multiple \glspl{du}. Although these works showcase the advantages of using accelerators in improving the performance of \gls{ran} deployments, they do not focus on automating the configurations of the infrastructure, and are limited to simple and non-scalable deployments mostly based on bare-metal implementations.

From the orchestration perspective, 
SoftRAN~\cite{softran} introduces a software-defined model that replaces the traditional distributed control plane with a logically centralized controller, which improves the global network optimization, management, scalability, and coordination. 
OrchestRAN~\cite{orchestran} proposes optimized allocations of resources to \glspl{gnb} varying from edge to cloud.
ATHENA~\cite{athena}, instead, is a cloud-native, multi-x network management and orchestration framework for the automation of network workloads lifecycle. It has a declarative, intent-based, and multi-vendor-compatible software-defined architecture that aims at improving the network management flexibility. \cite{mcmanus2024cloudbasedfederationframeworkprototype}~proposes a cloud-based federation framework that automates testbed integration, resource sharing, and remote experimentation on Amazon Web Services. It supports heterogeneous testbeds, allowing seamless access to distributed research infrastructure.
Building on these efforts to automate and virtualize Open \gls{ran} infrastructure, recent research has begun to explore the role of \glspl{llm} in further simplifying and streamlining network configuration and management. 
Some works explore solutions to automatically produce network configurations
using language models~\cite{wei2025leveragingllmagentstranslating}.
Others focus on mobile radio network and \gls{5g} deployments, e.g.,~\cite{10685369}, where the authors analyze various applications for domain knowledge, code and network configuration generation.

Compared to these works, which focus on features such as acceleration, orchestration, automation, or \glspl{llm}, \framework provides 
a holistic approach that covers all aspects of deployment and testing of a cellular network. 
It is also the first solution to simplify cellular network operations and management by enabling the intent-based deployment of network workloads and tests on a heterogeneous infrastructure with accelerators, different \gls{cpu} architectures, and multi-vendor 5G software and radio devices.

\section{\framework Cloud-Native Design}
\label{sec:e2e-telco-cloud}%

\begin{figure*}[ht!]
  \centering
  \includegraphics[width=.65\linewidth]{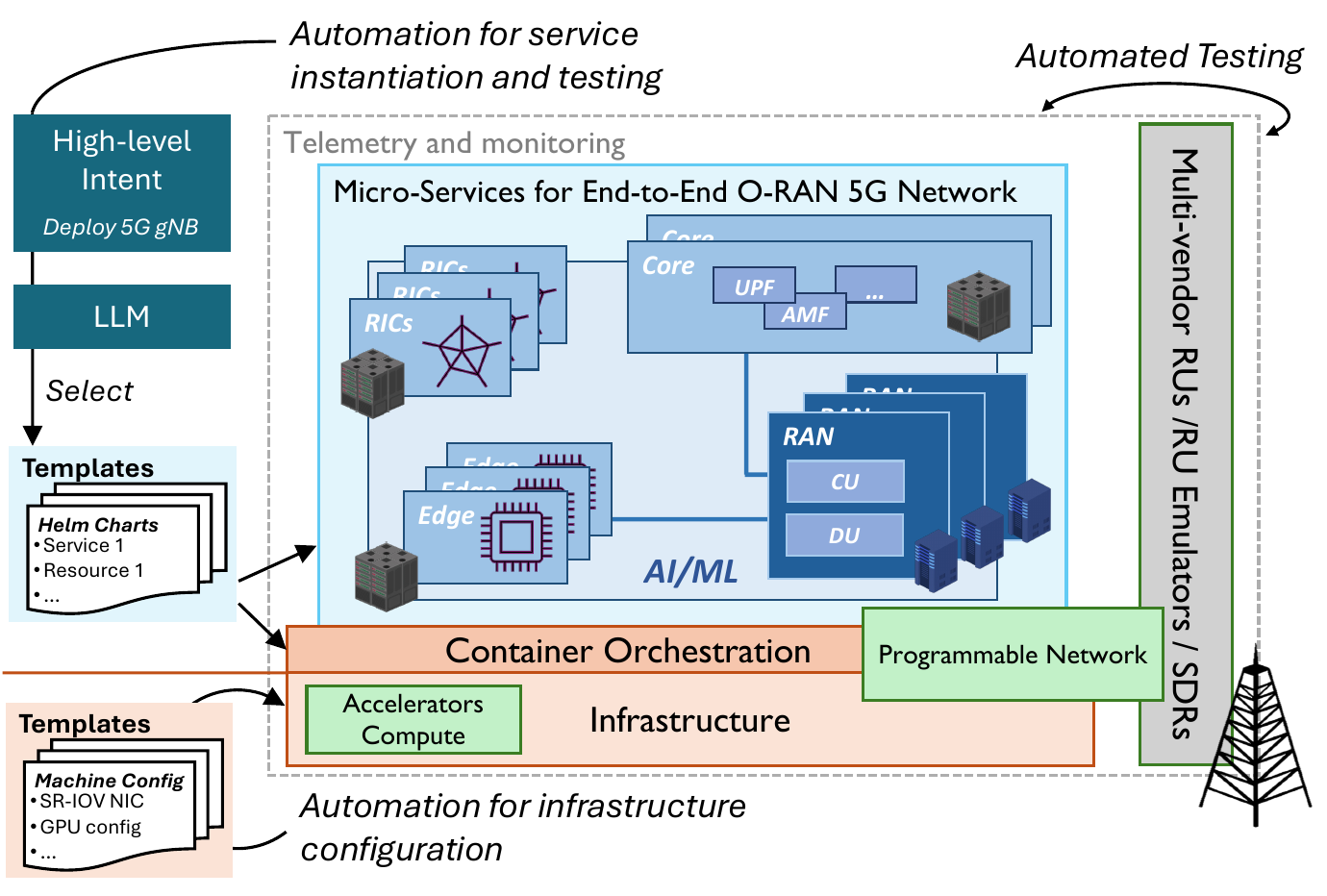}
  \vspace{-.3cm}
  \caption{\framework architecture and foundational components.}
  \vspace{-.3cm}
  \label{fig:cloud}
\end{figure*}

In this section, we discuss the design of \framework, focusing on the foundational principles that we leveraged to introduce zero-touch automation in \gls{e2e} software-driven cellular networks (Figure~\ref{fig:abstraction}). Specifically, we review how virtualization, \rifc, disaggregation and micro-services, orchestration, intent-based configuration, and observability are leveraged in \framework, whose high-level architecture is shown in Figure~\ref{fig:cloud}.

The systems and services that are required for \gls{5g} networks combine (i) elastic workloads, for core network, orchestration, and management-related services; and (ii) the \gls{ran}, which has stringent performance guarantees requirements to run complex \gls{dsp}. 
Due to their nature, elastic workloads can be placed directly on the infrastructure (e.g., at the edge, or in the cloud) with the caveat that connectivity and latency requirements are satisfied (e.g., a core network may run in cloud data centers such as AWS or Azure as long as the latency to reach the cloud is not too high). As shown in Figure~\ref{fig:cloud} (light blue), these include components for the core network, \glspl{ric}, and various non-latency-sensitive \gls{ml} applications.
In contrast, \gls{ran} workloads (dark blue in Figure~\ref{fig:cloud}) need to be distributed to edge or cell site locations, close to the end users, to satisfy the high data rates and predictable low latency requirements needed by \gls{ran} functions \cite{oran2024cus}.
For instance, the deployment of a \gls{gnb}---which comprises the disaggregated elements \gls{cu}, \gls{du}, and \gls{ru}---requires that \gls{du} and \gls{ru} are in close proximity to minimize latency over the front-haul interface carrying high-capacity traffic related to I/Q streams.
Additionally, \glspl{du} at cell sites might also need to host low-latency control and sensing applications such as dApps~\cite{doro2022dapps, dapp-ngrg-report, lacava2025dappsenablingrealtimeaibased}. These might perform spectrum sensing, beam management, anomaly detection or channel estimation, which need direct access to real-time I/Q sample streams, which calls for edge deployments rather than cloud ones due to the high-capacity bandwidth requirement to transmit I/Q streams.

Figure~\ref{fig:cloud} shows how \framework leverages a fast and programmable network to implement the fronthaul interface between the DU and multiple options as radios, including commercial \glspl{ru}, \gls{ru} emulators, and software-defined radios.
To holistically manage and optimize such a diverse architecture, \framework implements automated workflows for infrastructure configuration, \gls{e2e} deployment, and testing, based on the following design principles.

\subsection{Abstracting Compute, Networking, and Acceleration}
\label{sec:cluster}%
As we discuss in Section~\ref{sec:integration_details}, \framework is built on top of an infrastructure with a diverse set of compute, networking, and acceleration resources. To manage this complexity and diversity, we design \framework to abstract each individual component of the infrastructure via virtualization, thus creating a homogeneous software layer that hides the details of the  underlying infrastructure to simplify deployment and control. 
At the same time, we configure the infrastructure nodes and software (e.g., through proper kernel profiles) so that the virtualization overhead does not compromise the performance compared to bare metal, non-virtualized setups, as shown in Section~\ref{sec:performance-arc}. 
We leverage Podman as the base container technology for OpenShift, and rely on \gls{sriov} to virtualize the \gls{nic} and enable multiplexing of different traffic flows on the same interface. \glspl{gpu} are presented to the workloads through the NVIDIA Docker plugin for \glspl{gpu}, which makes sure that containers can fully access their compute capabilities. 
Finally, to account for different \gls{cpu} architectures and guarantee optimized performance of \gls{cpu} operations, we avoid translating instructions and develop a native, multi-architecture build pipeline that accounts for both ARM and x86.

Further, the infrastructure is organized as a cluster, i.e., an abstraction where compute nodes work under a centralized control plane (e.g., an orchestrator as in Section~\ref{sec:orchestration}), rather than as isolated servers. Combined with virtualization, this approach simplifies deployment and management by replicating configurations across nodes, supports heterogeneous computing, and enables on-demand reconfiguration. Applications can migrate between equivalent nodes for load balancing.
Moreover, it allows for a logical separation of data and application logic, e.g., to provide resilient and dynamic \gls{ran} services with storage for state and configurations that persists the life cycle of individual services.

\subsection{Declarative Approach for \rifc}
\label{sec:declarative}%

As discussed above, the \framework cluster combines a heterogeneous set of compute, acceleration, and networking components, and additional abstractions on top of it. The system needs to go through multiple stages for the configuration: (i) at pre-deployment, for planning (i.e., day 0); (ii) at deployment, i.e., whenever new devices or services need to be added or for updates (i.e., day 1); and (iii) for post-deployment, to manage the complete life-cycle (day 2). 
Manually configuring the cluster infrastructure and software is a task that is time consuming, prone to errors, and lacking verifiability and accountability. Misconfigurations can cause outages and instability in the system, and a non-systematic approach to configuring \framework may lead to delays in troubleshooting and detecting root causes for failures.

Therefore, we introduce a \rifc declarative approach for \framework, as shown through the template flows in Figure~\ref{fig:cloud}. We define the status and configuration of the system through templates and configuration files, with key/value pairs describing how each hardware and software component of \framework needs to be managed (e.g., specifying how many cores are reserved or isolated, \gls{gpu} configuration). Specific examples are provided in Section~\ref{sec:integration_details}. We implement a \emph{GitOps} workflow to version, track, and deploy such configurations. A central \texttt{git} server hosts the template files, which include machine configurations, Dockerfiles, Helm charts, and other text-based templates. \cicd pipelines automatically synchronize the repository on the \texttt{git} server with the hardware and services on the cluster. 
This guarantees that updates are automatically synchronized and aligned across production and the versioning server, and enforces a single and specific workflow to propagate changes and updates.  

The \cicd for the declarative approach is implemented through two main components, \textit{ArgoCD} and \textit{Tekton}.
ArgoCD is used to synchronize configurations between the \texttt{git} server (e.g., GitHub) and the \framework cluster. This also guarantees a stateless cluster, which can be deployed (or re-deployed) in few simple steps, compared to a long set of tedious manual configurations. 
Tekton is an open-source framework for designing and running \cicd.
Tekton pipelines consist of a declarative set of tasks executed one after the other, configured from sets of parameters passed as input. The parameters can be provided through default values as part of the pipeline definition, customized to ingest the output of other processes (as we discuss in Section~\ref{sec:aut-llm}), or manually overridden. Tekton can be accessed through \glspl{api}, making it easier to deploy applications on demand from within and outside the cluster.

\subsection{Disaggregation, \glspl{api}, and Micro-Services}

The software that constitutes \framework components is deployed as a set of micro-services which are atomic---yet connected---units deployed as pods on the cluster. A pod is a set of one or more Docker containers that provide functionalities to fully express a micro-service. \framework micro-services include the actual applications for the cluster, i.e., \gls{ran}, core network, \glspl{ric}, and edge services, as shown at the center of Figure~\ref{fig:cloud}, and the software that supports the cluster and automation itself. In this sense, there exist a unified workflow and management procedures for both classes of services, simplifying the overall design of the system.

As we discuss in Section~\ref{sec:integration_details}, a specific effort has been put into the design of the micro-services to (i) identify the degree of disaggregation that enables flexibility, automation, and scaling without compromising performance (e.g., whether different micro-services can be used for L1 and L2/L3 in a \gls{du}, or whether to split \gls{du} and \gls{cu} in different pods, among others); and to (ii) separate state and logic as much as possible. The latter allows the deployment of lightweight micro-services that embed the application logic but not complex data structures, which are instead on a permanent storage layer. This provides redundancy through disk replication via \gls{raid} and exposes its resources to the micro-services, or pods, so that they do not have to replicate state whenever they are deployed or perform complex operations on tear-down to store and manage the state. 

The disaggregation is based on a functional split. For the \gls{5g} \gls{e2e} application, it is aligned to \gls{3gpp} and O-RAN specifications, which include a \gls{sba} for the core network and a \gls{gnb} split into \gls{cu} and \gls{du} (software-based) and \gls{ru}. 
Furthermore, the \glspl{ric} are also deployed as set of micro-services, which can be extended by onboarding custom logic (xApps, rApps).
Interfaces for cellular network components are typically defined by standards or technical specifications from \gls{3gpp}, facilitating functionality over IP networks. In contrast, services such as automation and cluster functionalities are functionally split without strict adherence to specific standards. Their interfaces are implemented via application-level endpoints or APIs, often based on frameworks like Flask. Detailed API designs for various micro-services (e.g., automated testing or intent-based deployment) are discussed in Section~\ref{sec:automated-workflows}.

Similarly to the infrastructure, micro-services are also defined through a declarative approach, with Helm charts defining a set of unit elements to deploy (e.g., containers in a pod, the associated storage, networking capabilities, among others) and Dockerfiles specifying the features within a specific container (e.g., what software is used to execute \gls{du} functionalities). As discussed in Section~\ref{sec:integration_details}, we have designed an automated container build process targeting multiple architectures, as well as continuous deployment solutions, all based on Tekton pipelines. 

\subsection{Orchestration}
\label{sec:orchestration}%
The micro-services lifecycle is managed through an orchestrator, which takes care of handling the complexity associated to matching resources, micro-services, configurations, and cluster capabilities. The orchestrator takes care of micro-services deployment, scaling, networking, and lifecycle management, based on declarative input provided through the \cicd approach discussed in Section~\ref{sec:declarative}. Once the desired state of the system is defined (e.g., the number of replicas or specific configurations), the system continuously self-heals to maintain that state. This is done automatically, solving problems related to system complexity, ensuring high availability through automatic failover and scaling, and optimizing resource usage. 
The orchestrator controls operations over the infrastructure cluster, as defined in Section~\ref{sec:cluster}, matching micro-services to available resources. For example, it can ensure that a \gls{du} requiring \gls{gpu} acceleration is instantiated on a compute node with an available \gls{gpu} and \gls{nic}.

In our setup, and as described in Section~\ref{sec:integration_details}, we use Red Hat OpenShift, a commercial version of Kubernetes, as the orchestrator engine for our micro-services, which we tune, instrument, and configure to support the variety of \gls{5g} workloads provided by \framework.
Compared to other orchestration and virtualization tools, OpenShift provides superior flexibility and scalability. 
For instance, Ubuntu MaaS supports automated infrastructure deployment but lacks a container-based, parallel, and multi-node design. 
Ansible Playbooks enable automated deployments but do not offer the interactive and hybrid experimentation environment of \framework. 
Docker-based approaches face scaling issues with technologies such as \gls{sriov} and \gls{mig}, due to limited intent-based resource management. 
In general, most alternatives manage independent nodes rather than unified clusters. 
By contrast, Kubernetes natively supports multi-node orchestration, and OpenShift extends it with integrated automation tools like Tekton, delivering a production-ready solution.

The orchestrator also provides additional tools that can be used to manage the system. \framework leverages (i) namespaces, i.e., logical partitions to organize micro-services across, for example, application domains (e.g., pods for a core network, a \gls{ric}, the \gls{ran}) or tenants (e.g., different operators sharing the same infrastructure); and (ii) advanced networking capabilities, which enable automated service discovery and the establishment of complex network overlays across micro-services (e.g., dynamically establishing routes between core network micro-services and new \glspl{cu} instantiated through automation).

The orchestrator, finally, introduces a zero-touch approach into the network and system configuration, behaving as an intent-driven platform for automated deployment and testing. Although the automation capabilities of \framework can enable fully autonomous zero-touch operations, in this paper we focus on their use for streamlining and automating deployments, test generation and execution.

\subsection{High-Level Intents to Represent Network Status}
\label{sec:intents}%
While automation, \cicd, orchestration, and declarative approaches simplify the management of a \gls{5g} network from a system perspective, they still represent complex tools to use for a variety of end users that could be interested in managing and operating such networks. Consider, for example, private \gls{5g} deployments, where the enterprise \gls{it} team comes with limited knowledge on radio systems: expressing a rich configuration for the system components may be challenging, even if it is then automatically deployed and applied to the network. Similarly, teams with expertise on radio systems may have more limited knowledge on configuring orchestrators and micro-services. 

Therefore, we design \framework with a mechanism to provide high-level input for the system configuration, which is then translated into actionable templates and pipeline triggers that are automatically applied to the system. As shown in the top left part of Figure~\ref{fig:cloud}, users can express high-level requests to instantiate services or tests (e.g., \emph{``deploy a 5G base station''}, which are parsed through an \gls{llm} matching the intent with templates to be applied to the system. We provide details on the implementation of this mechanism in Section~\ref{sec:automated-workflows}.

\subsection{Observability}
\label{sec:monitoring-tools}%

Finally, once the system is configured, deployed, and operational, the \framework design also embeds procedures to automatically monitor the system status and coordinate with the orchestrator to maintain the desired system state. \framework leverages Prometheus~\cite{prometheus}, OpenShift APIs, and Observium~\cite{observium} to track application performance, resource consumption, and network status in real time. These tools work together to capture and visualize key metrics, e.g., CPU load, memory usage, and network traffic, and enable identification of anomalies, resource management, and continuous improvements to the overall system reliability and efficiency. 

\section{\framework System Components}
\label{sec:integration_details}%

The \framework framework runs on a multi-architecture Red Hat OpenShift cluster capable of virtualizing and streamlining the deployment of application workloads on a generic infrastructure, which can optionally interface with specialized hardware.
These workloads include \gls{ran}, core network components, \glspl{ric}, dApps, xApps, and rApps, \gls{ai} services and functions, as well as those that handle the automated deployment of a fully compliant \gls{5g} network managed by \oran.
Unlike bare metal deployments, \framework leverages the virtualization capabilities offered by OpenShift for enhanced flexibility and dynamic reconfiguration of the elements deployed on the \gls{5g} network.

In this section, we will describe the core components that enable the automation performed by \framework.
Section~\ref{sec:automation-cluster} describes the OpenShift cluster leveraged by \framework, with operators and configuration profiles detailed in Sections~\ref{sec:operators} and~\ref{sec:profiles}, respectively.
Section~\ref{sec:integrating-ran-deployments} describes how to integrate and deploy \gls{ran} workloads using the O-RAN 7.2 split, with and without a layer-1 accelerator, and with \glspl{sdr} based on a 8.1 split.
Finally, Section~\ref{sec:managing-images-and-building} details how to manage and build images for heterogeneous architectures, while Section~\ref{sec:additional-software-components} reviews additional software components of the cluster.

\subsection{The Automation Cluster}
\label{sec:automation-cluster}%

\begin{figure}[ht]   
  \centering
  \includegraphics[width=0.9\columnwidth]{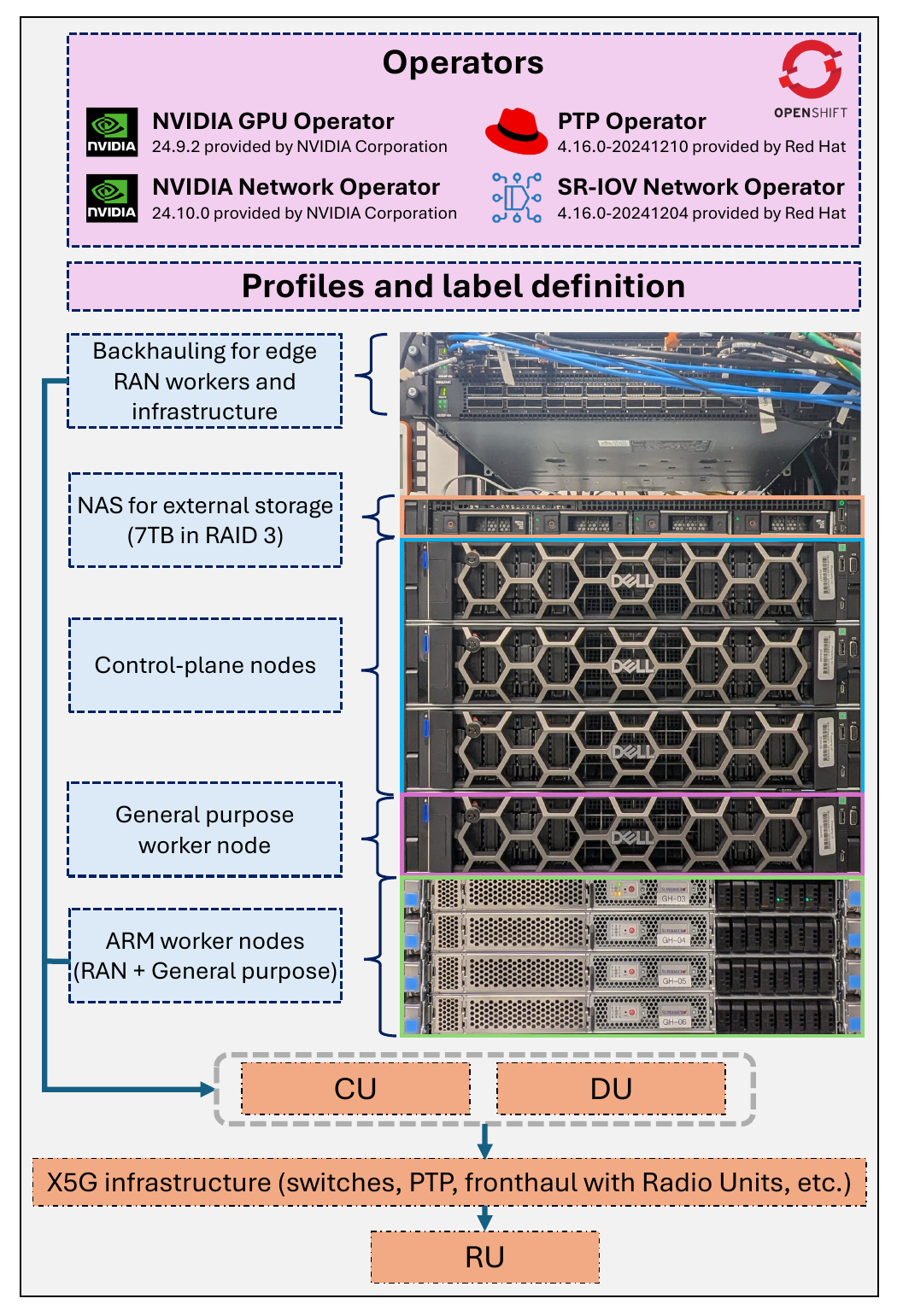}
  \caption{An overview of the \framework cluster.}
  \label{fig:cluster}
\end{figure}

The \framework architecture is depicted at a high level in Figure~\ref{fig:cluster}.
\framework is hosted in an OpenShift cluster composed of 13 nodes with heterogeneous computing hardware (summarized in Table~\ref{tab:list-nodes}) and a dedicated \gls{nas}.
\begin{table*}[h]
    \centering
    \caption{Node architectures being integrated in the AutoRAN framework.}
    \begin{tabular}{llll}
        \toprule
        Model & CPU & GPU & NIC \\
        \midrule
        Dell R760 & Intel Xeon 8462Y+ & NVIDIA L40S (40GB) & Broadcom \\
        Microway EPYC & AMD EPYC 7262 & N.A. & Mellanox ConnectX-6\\
        Gigabyte E251 & Intel Xeon 6240R & NVIDIA A100 (40/80 GB) & Mellanox ConnectX-6\\
        Supermicro ARS-111GL-NHR & Grace CPU & NVIDIA H100 Tensor Core GPU & Mellanox ConnectX-7 \\
        \bottomrule
    \end{tabular}
    \label{tab:list-nodes}
\end{table*}
The backhauling and networking infrastructure 
leverages \gls{sdn} switches with up to $100$\:Gbps connectivity and \gls{ptp} synchronization through a rooftop-mounted GPS antenna and local Qulsar clocks~\cite{villa2024x5g}.

Compute nodes are differentiated into control-plane and worker nodes. The former are used to host automation, control and monitoring functionalities of the cluster, as well as generic core network, \gls{ric}, and \gls{ai} workloads. The latter are optimized to host \gls{ran} workloads and services via low-latency configurations and \gls{gpu} acceleration capabilities. 
However, concurrently supporting \gls{ran} and generic cloud computing workloads comes with specific requirements on the underlying infrastructure. As detailed in Table~\ref{tab:list-nodes}, we select high-performance servers with different architectures (x86 and ARM) and acceleration capabilities (e.g., the Grace Hopper MGX GH200 ARM nodes used to run the GPU-accelerated NVIDIA ARC solution) to meet the stringent \gls{ran} requirements.
Key considerations also include maximizing physical CPU cores, as \gls{ran} workloads benefit from disabled hyperthreading functionalities~\cite{oai-hyperthreading}, and require isolated cores for optimal performance.

Servers connect to the cluster via $10$\:Gbps connections to the backhaul switch,
and to the fronthaul interface and \glspl{ru} via dedicated $100$\:Gbps connections to satisfy fronthaul traffic requirements (e.g., each $4\!\times\!4$ MIMO \gls{ru} with standard 9-bit compression requires a $10$\:Gbps connection). \framework also consolidates multiple \glspl{du} on each node, thus calling for a $100$\:Gbps \gls{nic}.
A \gls{ptp}-compatible \gls{nic} is also needed for carrying fronthaul traffic while maintaining precise synchronization through \gls{ptp} timestamping.
Finally, \framework also includes a persistent storage via a dedicated \gls{nas}. We separate \textit{computing} from \textit{storage} resources for resilience, thus obtaining a \textbf{stateless} cluster.
Specifically, the \gls{nas} used by \framework includes $12$\:TB of storage in RAID~3 using the open-source TrueNAS solution~\cite{truenas} to share volumes over the NFS protocol. The RAID~3 redundancy level implements a replica of the data over three separate disks, where two are used for data, while the third one is used for parity, which makes the usable disk space $8$\:TB.

\subsection{Operators for Automatic Configuration}
\label{sec:operators}

We leverage \textit{operators} for the configuration of \framework OpenShift nodes. These provide an high-level declarative approach for deterministic configuration of nodes of the cluster. The most relevant operators used in \framework are summarized in Table~\ref{tab:operators} and detailed in the remaining of this section.

\begin{table}[ht]
\centering
\caption{List of OpenShift operators.}
\label{tab:operators}
\begin{tabularx}{\columnwidth}{
        >{\raggedright\arraybackslash\hsize=0.75\hsize}X
        >{\raggedright\arraybackslash\hsize=1.25\hsize}X }
\toprule
Name & Function \\
\midrule
Node Feature Discovery      & Exports hardware and software features of nodes and label nodes    \\
NVIDIA GPU Operator         & Installs NVIDIA GPU drivers on all nodes and exposes various capabilities   \\
NVIDIA Network Operator     & Upgrades and configures NVIDIA NIC firmware  \\ 
SR-IOV operator             & Splits the NIC into partitions and instantiates networks  \\ 
PTP operator                & Provides ns clock accuracy to the pods on the node  \\ 
\bottomrule
\end{tabularx}%
\end{table}

\textbf{Node Feature Discovery.}~Since cloud-based deployments may run on very different hardware, the \gls{nfd} automatically expose the features of the underlying physical infrastructure (e.g., CPU model and architecture, number of cores, amount of memory, GPU availability and type, among others) to applications and workload using labels. 
These labels allow the cluster to perform targeted deployments, either on a particular set of nodes or on the specific node. This ensures that components, operators, and deployments are instantiated on the nodes with the required resources.

\textbf{PTP Operator.}~\gls{ran} systems require precise timing synchronization across their components. In general, clock synchronization is achieved through the \gls{ntp}, which is able to provide \textit{millisecond} accuracy over \gls{lan} setups. However, the level of accuracy delivered by \gls{ntp} is not sufficient to support \gls{ran} applications, where the Open Fronthaul interface requires \textit{nanosecond} accuracy.
We therefore use \gls{ptp}, a protocol that shares a synchronization signal over Ethernet, to sync all nodes and radios to a common clock source---a Qulsar clock with a GPS input in our case.
The Linux implementation of the protocol uses two pieces of software, namely \texttt{ptp4l} and \texttt{phc2sys}, which are installed via the OpenShift \gls{ptp} operator. 
The operator also gives the possibility of increasing the priority of these processes, allowing one to use them on shared CPUs rather than on a dedicated CPU per process, saving one extra core for other workloads.
Listing~\ref{lst:ptpexample} shows the \gls{ptp} configuration adapted for the Grace Hopper server.
\begin{lstlisting}[float=ht,floatplacement=h,language=json,style=mystyle-json, 
caption={Example of a PTP profile on a Grace Hopper node.}, 
label={lst:ptpexample}]
apiVersion: ptp.openshift.io/v1
kind: PtpConfig
metadata:
  name: ptp-gh
  namespace: openshift-ptp
spec:
  profile:
    - interface: enp1s0f0np0
      name: ptp-gh
      phc2sysOpts: '-a -r -r -n 24'
      ptp4lConf: |
        [global]
        dataset_comparison              G.8275.x
        G.8275.defaultDS.localPriority  128
        maxStepsRemoved                 255
        logAnnounceInterval             -3
        logSyncInterval                 -4
        logMinDelayReqInterval          -4
        G.8275.portDS.localPriority     128
        network_transport               L2
        domainNumber                    24
        tx_timestamp_timeout            30
        slaveOnly 1
        
        clock_servo pi
        step_threshold 1.0
        egressLatency 28
        pi_proportional_const 4.65
        pi_integral_const 0.1

        [enp1s0f0np0]
        announceReceiptTimeout 3
        delay_mechanism E2E
        network_transport L2
      ptpClockThreshold:
        holdOverTimeout: 5
        maxOffsetThreshold: 50
        minOffsetThreshold: -50
      ptpSchedulingPolicy: SCHED_FIFO
      ptpSchedulingPriority: 65
  recommend:
    - match:
        - nodeLabel: node-role.kubernetes.io/worker-gh
      priority: 4
      profile: ptp-gh
\end{lstlisting}

\textbf{NVIDIA Operators.}~These include the NVIDIA GPU Operator and the NVIDIA Network Operator.
The NVIDIA GPU Operator is in charge of installing and maintaining NVIDIA GPU drivers. GPUs are used in \framework for both \gls{ai} workloads (e.g., \glspl{llm}, training and testing of models) and for the NVIDIA ARC deployment~\cite{villa2024x5g}, which uses GPUs to accelerate the \gls{du}-low operations \gls{5g} \glspl{gnb}.
Specifically, NVIDIA ARC requires the Open Kernel Modules instead of the proprietary drivers. We also provision GDRCopy~\cite{gdrcopy} in the NVIDIA operator so that it is automatically deployed on all nodes and can directly access the GPU memory. This is a low-latency GPU memory copy library based on NVIDIA GPUDirect RDMA technology, that creates the CPU mapping of GPU memory.

The NVIDIA Network Operator is used to enable intercommunication between GPU and \gls{nic}. It is used to access the firmware of the \gls{nic}, for example, to enable higher timing accuracy and specific \gls{qos}.

\textbf{SR-IOV Operator.}~This operator is in charge of creating virtual slices---or \textit{virtual functions}---of the physical \gls{nic} interfaces and to expose them as objects inside OpenShift.
In OpenShift and Kubernetes deployments, these are created using configuration files called \texttt{\gls{sriov} Networks Node Attachment}.
After splitting the interfaces, OpenShift implements \texttt{Network attachments} to specify the IP addressing of the network and the VLAN tag. As it will be discussed in Section~\ref{sec:integrating-ran-deployments}, NVIDIA ARC deployments require interfaces to be passed as untagged, whereas \gls{oai} requires the tag to be set on the fronthaul port.
This becomes key as different deployments need different settings on the physical port, and it enables parallelization of multiple workloads
on the same compute nodes. Additionally, without using \gls{sriov}, a \gls{nic} can only be accessed by a single container (e.g., a \gls{du}) at a time, whereas, with \gls{sriov}, it is possible to run multiple applications requiring access to same physical interface by connecting each of them to the virtual interfaces created by \gls{sriov}.
This is used, for instance to parallelize multiple NVIDIA ARC deployments by running the L1 over the virtualized \gls{sriov} interface, where we partition each \gls{nic} into 8 virtual interfaces. For this number of partitions, we do not observe a significant increase in latency compared to bare metal or Docker-based setups.
To the best of our knowledge, we are the first in following this approach with NVIDIA ARC.

An example of definition of networks over \gls{sriov} devices is shown in Listing~\ref{lst:sr-iov-example}. A \texttt{resourceName} exposes the underlying sets of virtual functions (8 in this case) created from the selected \gls{nic}. A network is then created on the virtualized \gls{nic} with parameters including VLAN tags, and IPs.

\begin{lstlisting}[float=ht,floatplacement=h,language=json,style=mystyle-json, 
caption={Example of a SR-IOV network policy and its network attachment.}, 
label={lst:sr-iov-example}]
apiVersion: sriovnetwork.openshift.io/v1
kind: SriovNetworkNodePolicy
metadata:
  name: gh-vf-sriov
  namespace: openshift-sriov-network-operator
spec:
  resourceName: vfgh
  nodeSelector:
    node-role.kubernetes.io/worker-gh: ""
  numVfs: 8
  mtu: 9216
  priority: 1
  nicSelector:
    vendor: "15b3"
    deviceID: "a2dc"
    rootDevices:
      - "0000:01:00.0"
  isRdma: true
  needVhostNet: true
  deviceType: netdevice
---
apiVersion: sriovnetwork.openshift.io/v1
kind: SriovNetwork
metadata:
  name: gh-vnf-net
  namespace: openshift-sriov-network-operator
spec:
  ipam: |
    ## any IP address configurations, if needed
  resourceName: vfgh
  networkNamespace: aerial
  # vlan: 2 # if a VLAN tag is needed
\end{lstlisting}

\subsection{Configuration Profiles}
\label{sec:profiles}%

Due to the stringent latency and processing requirements of \gls{ran} workloads, it is important to make sure that host machines use an appropriate kernel version and are configured to meet such requirements.
This includes setting isolated CPUs, HugePages and disabling energy saving functionalities that might put the processor in an idle state and decrease performance.
%
We use labels to automatically apply configurations to nodes upon their addition to the cluster. Nodes with the same physical hardware and configurations are grouped together in an \gls{mcp}. Then, for each \gls{mcp}, we apply the same set of configurations.

We use \texttt{PerfomanceProfile} objects to: (i)~configure the number of \textit{reserved} and \textit{isolated} cores; (ii)~enable HugePages to directly allocate blocks of memory ($1$~GB for x86 nodes, $512$~MB for ARM deployments); and (iii)~fine-tune kernel parameters related to interrupts and energy, e.g., disabling sleep states and offsetting the periodic clock ticks to mitigate jitter.
An example of a \texttt{PerfomanceProfile} for a Grace Hopper node is shown in Listing~\ref{lst:performance-profile-example}.
\begin{lstlisting}[float=ht,floatplacement=h,language=json,style=mystyle-json, 
caption={Example of performance profile for a Grace Hopper machine.}, 
label={lst:performance-profile-example}]
apiVersion: performance.openshift.io/v2
kind: PerformanceProfile
metadata:
  name: gh-performanceprofile
  annotations:
    performance.openshift.io/ignore-cgroups-version: "true"
    kubeletconfig.experimental: |
      ## various kubelet parameters for fine-tuned optimizations
spec:
  additionalKernelArgs:
    - "tsc=reliable"
    - "nohz_full=4-64"
    - "preempt=none"
    - "..."
  cpu:
    isolated: 4-64
    reserved: 0-3,65-71
  hugePages:
    defaultHugePagesSize: "512Mi"
    pages:
      - size: "512Mi"
        count: 48
  nodeSelector:
     node-role.kubernetes.io/worker-gh: ""
  machineConfigPoolSelector:
     machineconfiguration.openshift.io/role: worker-gh
  numa:
    topologyPolicy: "none"
  workloadHints:
    realTime: true
    highPowerConsumption: true
\end{lstlisting}
In addition, different \gls{ptp} profiles are also applied to each server, taking into account their hardware specifications and configuration parameters (for example, the name of the specific network interface that should receive the synchronization signal).

Profiles linked to a label are immediately applied to each new labeled node. Nodes update in a rolling strategy and, at the end of the update, their configurations are aligned, which eliminates the risk of stale drivers.
This flexibility in node labeling and zero-touch configuration is even more relevant for extendability and scalability of the framework.
Indeed, multiple profiles can be prepared in advance and applied to \glspl{mcp}. If a node---or a set of nodes---needs to be adapted for different workloads (from \gls{ran} to generic cloud computing or high-performance computing), it is only necessary to relabel the node so as to make it part of a different \gls{mcp} with different configurations, and the node gets automatically provisioned.

Finally, it is worth mentioning that differently from bare-metal or Kubernetes-based deployments, OpenShift only supports generic or real-time kernels (from no other sources than Red Hat itself), and does not support the low-latency ones generally recommended for \gls{gnb} protocol stacks.
Since at this time the NVIDIA L1 deployment requires drivers that are not available for Red Hat's real-time kernel, we adopt the generic Linux kernel on all the cluster nodes.
Experimental evidence in~\cite{bonati20235gct} also shows the instabilities of real-time kernels for \gls{ran} deployments, which results in poorer performance when compared to the generic one.

\subsection{Integrating RAN Deployments}
\label{sec:integrating-ran-deployments}

We validated a set of deployments that can be instantiated on the \framework infrastructure at this time. These deployments, shown in Table~\ref{tab:tested_configurations}, include deployments with both commercial radios with support for the O-RAN 7.2 split, \gls{sdr} (e.g., \glspl{usrp}), as well as CPU- and GPU-accelerated \gls{du} solutions.
\begin{table}[htb]
    \centering
    \footnotesize
    \setlength{\tabcolsep}{1pt}
    \caption{Protocol stack and OTA RU pairs validated on \framework.}
    \label{tab:tested_configurations}
    \begin{tabularx}{\columnwidth}{
        >{\raggedright\arraybackslash\hsize=0.75\hsize}X
        >{\raggedright\arraybackslash\hsize=0.75\hsize}X
        >{\raggedright\arraybackslash\hsize=1.5\hsize}X }
        \toprule
        gNB stack & Fronthaul/L1 & RUs \\
        \midrule
        NVIDIA ARC & GPU Acc. L1 & Foxconn RPQN (100 MHz) \\
        OAI & OSC FH & Foxconn RPQN (20/40/100 MHz) \\
        OAI & N/A & USRP x310/x410 \\
        srsRAN & N/A & USRP x310/x410 \\
        \bottomrule
    \end{tabularx}
\end{table}
We extend the USRP-based deployment proposed in~\cite{bonati20235gct, bonati2023neutran}. Pods are assigned one \gls{sriov} device, whose physical \gls{nic} is connected through a switch to the USRPs deployed in our laboratory environment. This kind of deployment resembles the \oran 8.1 split, since the entire logic of the disaggregated \gls{gnb} is focused on the \gls{cu}/\gls{du}. 
We, then, proceed to include the O-RAN 7.2 split-based deployments, focusing in particular on \gls{oai} and NVIDIA ARC.
This split divides the functionalities of a traditional \gls{bbu} into two parts: high-PHY processing remains in the \gls{du}, while low-PHY processing is moved to the \gls{ru}. 
This enables effective centralization of compute resources in the \gls{du}, while the \gls{ru} can be simplified and deployed closer to the antennas. 
It also supports robust signal processing capabilities and reduces latency by minimizing the fronthaul bandwidth requirements between \gls{du} and \gls{ru}. 

To support the 7.2 split, it is necessary to use high-performance and low-latency compute nodes that can sustain the high processing requirements of the Open Fronthaul and the low-\gls{du}.
\gls{oai} integrates an \gls{osc} library that provides support for the 7.2 split, while ARC uses an NVIDIA library for x86 and ARM architectures.\footnote{It is worth recalling that NVIDIA ARC also uses OAI for the \gls{du}-high and \gls{cu}.}
Moreover, both the OpenAirInterface and NVIDIA ARC platforms require \gls{dpdk}, a set of libraries that accelerate the processing of packets to support real-time operations.
For the 7.2 \gls{oai} deployment, we deploy a single pod containing \gls{oai} \gls{gnb}, which requires two tagged \gls{sriov} devices, one for the control plane and one for the user plane, each accelerated by \gls{dpdk}. 
\begin{figure}[h]
  \centering
  \includegraphics[width=0.8\columnwidth]{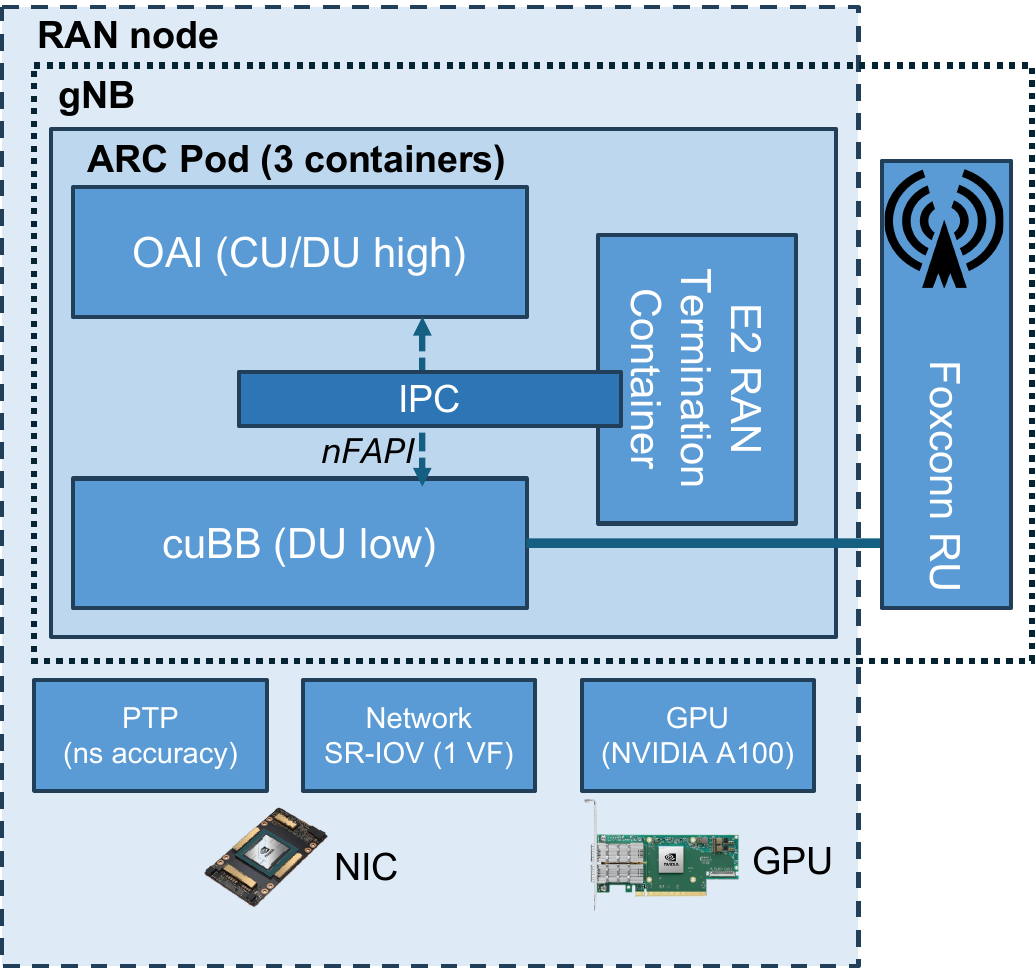}
  \caption{ARC deployment in \framework.}
  \label{fig:arc}
\end{figure}

Differently, the NVIDIA ARC setup requires a more complex deployment. Specifically, NVIDIA ARC usually runs as a combination of two containers, one for the L1 accelerator, called cuBB, and the other one for the L2, based on \gls{oai}.
The communication between L1 and L2 happens via \gls{ipc} shared between the two containers.
The original implementation from NVIDIA is based on Docker containers that access directly the NIC and the hardware using the Host Network Mode, that is without any layer of networking virtualization
Differently from this approach, we remove the host network requirement and we run NVIDIA ARC over virtualized \glspl{nic} using an untagged port since the L1 is in charge of adding the VLAN tag.
In OpenShift and Kubernetes, this can be achieved either via two pods, each with a single container, or a single pod with two containers.
We choose the latter deployment option (shown in Figure~\ref{fig:arc}, where we also add the E2 termination described in~\cite{villa2024x5g}) to support the instantiation of multiple NVIDIA ARC containers (e.g., to deploy multiple \glspl{gnb}) because both L1 and L2 are tightly coupled, and requires shared memory and networking.

All \glspl{gnb} based on the 7.2 split and without GPU accelerator share the same deployment structure. Therefore, when a new radio---real or emulated as RuSIM---gets integrated into \framework, the only modification required is in the template of the configuration file. The same applies, for both L1 and L2 configuration files, to the NVIDIA ARC deployment.
%

%

\setcounter{figure}{4}
\begin{figure}[th]   
  \centering
  \includegraphics[width=\columnwidth]{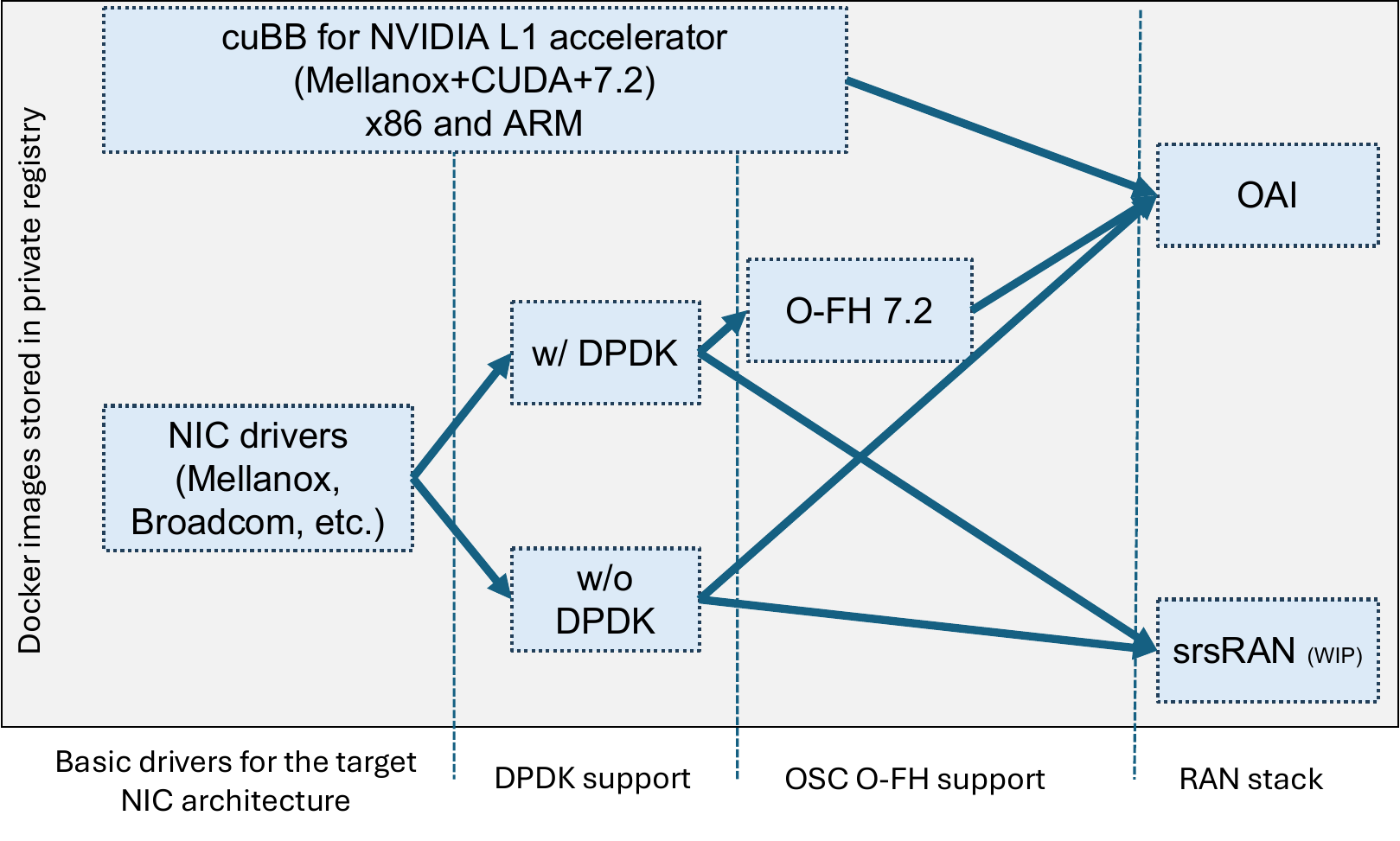}
  \caption{Chained building procedure.}
  \label{fig:build}
  \vspace{-.6cm}
\end{figure}

\setcounter{figure}{5}
\begin{figure*}[htbp!]
\setlength\abovecaptionskip{1pt}
    \centering
    \includegraphics[width=.9\textwidth,keepaspectratio]{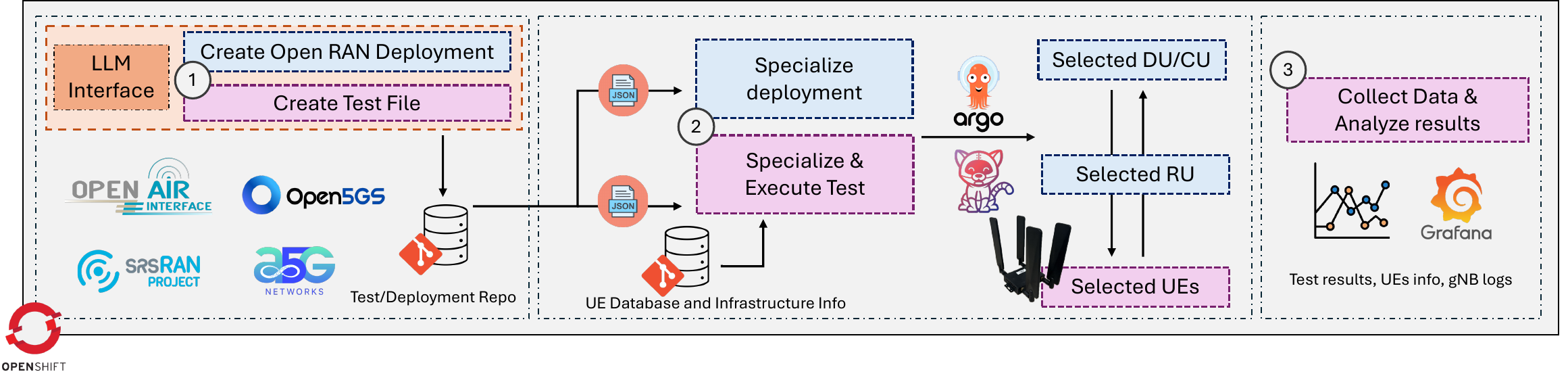}
    \caption{\gls{e2e} automated testing diagram. 
    }
    \label{fig:pipeline}
    \vspace{-.3cm}
\end{figure*}

\subsection{Container Image Building}
\label{sec:managing-images-and-building}

Another important aspect for a virtualized deployment is the building of container images.
Unlike cloud computing use cases, where images are generic, \gls{ran} deployments require instruction sets optimized for the specific CPU architecture and host family where the container will be instantiated.
A possible approach is to build the image for a specific architecture by specifying flags to the Dockerfiles (e.g., a \textit{cascadelake} architecture flag). This makes it possible to build images on any node in the cluster.
However, it requires different Dockerfiles for different deployment.
Another approach, which is the one we follow, consists in building the image on the target deployment node (e.g., if we want to run \gls{oai} on a Grace Hopper node, we build the image on a Grace Hopper directly).
In this way, we build images for each node without the need for different Dockerfiles.
As we will show later in Section~\ref{sec:coexistence}, thanks to the isolation between cores and processes, the build process does not affect the performance of the \gls{ran}.

To maximize image reuse and reduce build times, images are built in a chained manner, as shown in Figure~\ref{fig:build}.
Specifically, different \gls{gnb} images (e.g., \gls{oai}, srsRAN) and their subsequent versions (e.g., among different weekly tags in the case of \gls{oai}) might share common dependencies, e.g., in terms of \gls{nic} drivers, \gls{dpdk}, etc.
Instead of rebuilding all the dependencies for each new image, we maintain dedicated images with the required dependencies installed but no \gls{gnb} software.
We, then, use these images as a base to build the \gls{gnb} images, thus speeding up the build process at the cost of additional images (i.e., the dependencies images) stored on the registry.
Since \framework is deployed across heterogeneous server architectures (ARM and x86) and CPU models (e.g., Cascade Lake, Sapphire Rapids), we adopt an \gls{mcp}-specific build approach. 
To ensure that each \gls{ran} image fully leverages the capabilities of the hosting node, images are built directly on the nodes where the corresponding workloads will be executed.

\subsection{Additional Software Components}
\label{sec:additional-software-components}

In addition to the \gls{ran} workloads, a fully operational Open \gls{ran} requires several other elements. 
The core network is in charge of operations such as user authentication and the creation of \gls{pdu} sessions.
Among the various core networks available in the literature, we deployed Open5GS.
However, \framework is not bound to a specific core network, and we decided to focus on Open5GS due to of its high performance and ease of installations in Kubernetes-based deployments, where we deploy multiple independent replicas with different PLMNs for different tenants.

Another key element of an \oran deployment is the \gls{ric}, a component that enables observability and control of the network~\cite{polese2022understanding}.
Specifically, Near-Real-Time (Near-RT) control is achieved via applications called xApps deployed on the Near-RT \gls{ric}, while Non-RT control via rApps deployed on the Non-RT \gls{ric}
We deploy an \gls{osc} Near-RT \gls{ric} (``E'' release) and connect it to \gls{oai} via the E2 agent provided in~\cite{villa2024x5g,moro2023nfv} and publicly available as part of the OpenRAN Gym framework~\cite{bonati2023openrangympawr}.
Finally, the cluster also hosts a non-RT \gls{ric} to enable observability at larger time scales.

\section{\framework End-to-End Workflows} \label{sec:automated-workflows}
In this section, we focus on providing details on how to leverage such blocks to automate the deployment, management and testing of a private \gls{5g} network \gls{ota}. 
Specifically, we extend our previous work~\cite{bonati20235gct} to support a broader set of testing operations and technologies, deploy and test automatically different combinations of protocol stacks and \glspl{ru} (see Table~\ref{tab:tested_configurations}) in a repeatable manner, and analyze their performance under different conditions (e.g. number of users, target data rate, and arbitrary protocol configuration files). 
As we describe later, this is done through a set of pipelines that process high-level, user-specified requirements regarding the specific configuration to test (e.g. \gls{oai} with a certain \gls{ru} and MIMO configuration) and convert them into deployment and testing operations. These involve (i) instantiation of the \gls{5g} software containers as pods; (ii) initialization and attachment of \glspl{ue}; and (iii) data collection of relevant performance metrics to evaluate test success.
Additionally, such deployment and testing configurations can be automatically generated through \glspl{llm} starting from a high-level intent, and then actuated by the above pipelines, as discussed in Section~\ref{sec:aut-llm}.
To streamline and automate testing procedures and support remote execution of tests, our \glspl{ue} are Sierra Wireless EM9191 \gls{5g} modems connected to mini-PCs (e.g., Intel NUC, or Raspberry Pi). The \glspl{ue}'s mini-PCs are used as host machines that pilot the UE to perform attachment operations via AT commands embedded in the Qualcomm's chipset, generate traffic and collect data.

\textbf{\framework Worflows.} The workflows for the deployment and testing of Open \gls{ran} components are shown in Figure~\ref{fig:pipeline}. The blue blocks in the figure concern the instantiation of workload, while the pink ones carry out the functionalities related to testing.
The deployment workflow involves steps from the creation of generic deployment specifications, to their specialization to the hardware infrastructure and available resources.
Similarly, the test workflow concerns the creation of generic test specifications, their specialization to the testing infrastructure, the actual test execution, and data collection and analysis.
The steps for both deployment and testing workflows will be detailed in the remaining Sections~\ref{sec:deployment-workflow} and \ref{sec:testing-workflow} respectively.

\subsection{Intent-based Instantiation and Testing}
\setcounter{figure}{6}
\label{sec:aut-llm}%
\begin{figure*}
    \centering
    \includegraphics[width=.8\linewidth]{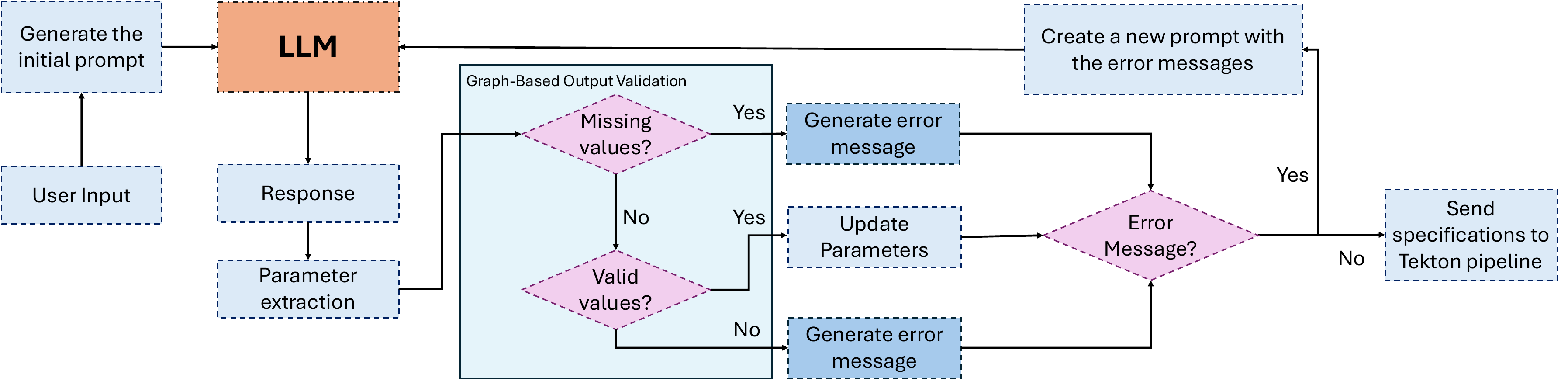}
    \caption{LLM agent for configuration extraction.}
    \vspace{-10pt}
    \label{fig:llm_diag}
\end{figure*}
In the following paragraphs, we describe how we leverage \framework to simplify and streamline network instantiation procedures and testing via intent-based deployment and testing, leveraging natural language processing. The goal of this feature is to eliminate complexity related to network configuration and allow even non-expert users to deploy and test a full-fledged network using only \gls{naturallanguage}. Examples of possible intents include queries such as \textit{``deploy a 5G \gls{gnb} with \gls{oai} and NVIDIA \gls{arc}''} or \textit{``perform a 15 Mbps iPerf test with 3~\glspl{ue}''}.
In order to process \gls{naturallanguage}, we opt for using \glspl{llm} as they provide capabilities to parse text, understand context and need, and convert them into structured and machine-understandable directives. Given the large availability of pre-trained models with excellent performance for a variety of tasks under different benchmarks, we opt for using a pre-trained \gls{llm} (Qwen 2.5) in an agentic setup, i.e., where the \gls{llm} makes calls to a set of tools (defined thereafter) in iterative loops.

Although appealing for their simplicity, \glspl{llm} are also well known for sometimes drifting away from initial instructions and producing hallucinations. In our case, this is particularly relevant, as any hallucination could result in incorrect network configuration that might produce wrong/misleading test results and even generate outages. To prevent this unwanted behavior, we design a set of procedures (depicted in Figure \ref{fig:llm_diag}) that constrain the output of the \gls{llm}, forcing it to output the correct network/test configurations.

Achieving this in practice is not trivial, and presents several challenges:

\begin{itemize}[leftmargin=*]
    \item First, it should be noted that not all the \gls{ran} components available in \framework are compatible with each other. For example, OAI supports the \gls{arc} L1, while srsRAN does not, meaning we need to prevent the \gls{llm} from trying to configure a srsRAN base station with \gls{arc}, which would result in a non-working deployment. It is therefore fundamental to ensure that the \gls{llm} only selects compatible components and configurations (e.g., \gls{oai} with \gls{arc}, plain \gls{oai}/srsRAN, as shown in Table~\ref{tab:tested_configurations}). For this reason, we store a compatibility graph, indicating which network components (DU, CU, RU, Core Network) are compatible with one another. Similarly, for each test type, a graph of the list of compatible parameters is fed to the \gls{llm} (for example, tests using different traffic generators, such as \gls{mgen} can include parameters such as the distribution of traffic, while iPerf tests do not). 
    In the workflow of Figure \ref{fig:llm_diag}, \revision{the compatibility graph is used in the process to validate the configurations generated by the LLM by making sure that those include elements that are compatible with one another.}
    When a new technology (\gls{cu}, \gls{du}, \gls{ru}) becomes available and integrated, the dependability graph is updated to reflect the compatibility between that specific element and the others available in \framework.
    
    \item Second, the \gls{llm} might output some text that drifts slightly from the anticipated output, causing potential parsing issues and undefined behaviors, such as outputting incorrect attribute values. We solve this issue by using the tool-calling feature of recent \glspl{llm}. With this feature, we can feed the \gls{llm} with a list of tools which it can explicitly call by outputting special tokens. In our case, those tools consist of a set of pre-defined Python functions which (i)~set the values of the network parameters in a dictionary (e.g., associate each possible parameter such as \gls{cu}/\gls{du}/\gls{ru}) with a value (in Figure \ref{fig:llm_diag}, this corresponds to the update of parameters); and (ii)~verify that the set values are correct, i.e., that they correspond to valid parameters which are compatible with one another (in the value validation step). 

    \item Another issue is that while most \glspl{llm} can follow instructions, the complexity of the network configuration task calls for multiple long instructions (this includes describing the use-case, the infrastructure, the compatibility graph, etc.). The result is that most models manage to follow some of the instructions, but rarely all of them at the same time. For this reason, instead of relying on a single prompting round, as shown in Figure~\ref{fig:llm_diag}, we resort to a looping mechanism in which the \gls{llm} is fed back with reports on misconfigurations and missing configuration fields. This enables the \gls{llm} to build the requested configuration iteratively, in a two phase manner where it first loops until all parameters have been populated with a correct value. Then, in the second phase, the validity is verified by ensuring that the configuration matches the compatibility graph.
\end{itemize}

Once we obtain the correct configuration (\textit{i.e.}, when there is no more error message in Figure \ref{fig:llm_diag}), we convert it into \texttt{json} format and send it to the correct pipeline (either the testing pipeline or the deployment pipeline) for execution on the cluster. This process is essential as it ensures we map intents to actionable network deployments, making sure that the selected software and hardware elements can interact with each other and can effectively deliver network services to users.

\subsection{Deployment Workflow}
\label{sec:deployment-workflow}

The deployment workflow is in charge of automatically instantiating the specified workloads (e.g., \gls{oai} \gls{gnb}, core network, \gls{ric}) on the virtualized infrastructure.
The main steps of this workflow, shown with blue boxes in Figure~\ref{fig:pipeline}, involve: (1)~creating the Open \gls{ran} deployment file; and (2)~specializing this to the specific hardware infrastructure abstracted by OpenShift (e.g., selecting specific \glspl{cu}/\glspl{du} and \glspl{ru}).
The deployment file is used to specify the workloads to instantiate on the physical infrastructure, e.g., an \gls{oai} \gls{cu}/\gls{du} with Foxconn \gls{ru} (see Table~\ref{tab:tested_configurations} for the list of possible combinations).
An example of this file (and possible output of the \gls{llm}) is shown in Listing~\ref{lst:deployment-example}, where the core network (marked as \texttt{core\_network} in the listing) is set to Open5GS, the protocol stack to \gls{oai} for \gls{cu}/\gls{du}-high (\texttt{cu} and \texttt{du-high}) with NVIDIA \gls{arc} for \gls{du}-low (\texttt{du-low}), and \gls{ru} to Foxconn (\texttt{ru}). 
\begin{lstlisting}[float=ht,floatplacement=h,language=json,style=mystyle-json, 
caption={Example deployment file with the components and configurations be tested.}, 
label={lst:deployment-example}]
{
  "network_scenario": {
    "id": 1,
    "core_network": {
      "name": "open5gs"
    },
    "cu": {
      "name": "oai",
      "config_file": "oai_100_3750_2x2.conf"
    },
    "du-high": {
      "name": "oai",
      "config_file": null
    },
    "du-low": {
      "name": "cubb",
      "config_file": "cubb_100_2x2.yaml"
    },
    "ru": {
      "name": "foxconn",
      "location": 660,
      "config_file": null
    }
  }
}
\end{lstlisting}

After building the deployment file, we leverage Tekton pipelines to specialize the parameters therein to specific infrastructure components (e.g., the selected \texttt{ru} is converted in the MAC address of the \gls{ru}), and to execute the actual deployment of the virtualized components.
This is done via pipelines synchronized on the OpenShift cluster via the ArgoCD framework from a version-controlled git repository (see Section~\ref{sec:e2e-telco-cloud}).

Each pipeline consists of multiple \textit{tasks} that need to be executed to successfully run the pipeline.
These tasks are associated to operations that span from sending the intent of the user to the \gls{llm}, to converting the output of the latter into a deployment object specialized to the physical infrastructure, to configuring and instantiating the gNB components (e.g., \glspl{cu} and \glspl{du}) and connecting them to the radio devices used for the test (e.g., either USRP \glspl{sdr} or commercial \glspl{ru}) and to the core network.
After instantiating the specified components in the form of virtualized containers, the gNB software is started via a call to Flask \glspl{api} that they expose.
\revision{For NVIDIA-ARC deployments, the \framework framework performs a single \gls{api} call to start the workload. Specifically, an \gls{api} endpoint is exposed by the \gls{cu}/\gls{du} container. This, then, applies the correct configuration for the deployment and, using an internal connection, starts the \gls{du}-low container.}
Thus, by leveraging OpenShift’s built-in virtualization and parallelization capabilities, \framework can execute multiple experiments concurrently, provided sufficient hardware resources are available and there is no interference between the
\glspl{gnb}.

\subsection{Testing Workflow}
\label{sec:testing-workflow}

The testing workflow is in charge of automatically executing performance tests leveraging the components instantiated by the deployment workflow (see Section~\ref{sec:deployment-workflow}).
Its main steps, shown in pink boxes in Figure~\ref{fig:pipeline}, concern: (1)~creating a test file; (2)~specializing it to the components under testing and running the actual test; and (3)~collecting and analyzing test data.
The test file, an example of which is shown in Listing~\ref{lst:traffic test-example}, specifies various parameters for each \gls{ue}, such as the network slice the \gls{ue} is assigned to, the type of test to run (e.g., iPerf, \gls{mgen}), target data rate, and test duration.

\begin{lstlisting}[float=ht,floatplacement=h,language=json,style=mystyle-json, 
caption={Example of iPerf downlink test file.}, 
label={lst:traffic test-example}]
{
  "network_scenario": {
    "id": 1,
    "ue_specification": [
      {
        "slice_id": 1,
        "test_type": "iperf",
        "bandwidth_mbps": 25,
        "duration": 60,
        "protocol": "udp",
        "reverse": true,
        "json_output": true,
        "server_hostname": "http://server.automation.otic.open6g.net",
        "server_port": 32201
      }
    ]
  }
}
\end{lstlisting}

The created test file is then fed to a Tekton pipeline, which activates the specified \glspl{ue} for them to connect to the deployed components (see Section~\ref{sec:deployment-workflow}).
Information on \glspl{ue} and their functionalities is stored in a \textit{\gls{ue} database}, an example of which is shown in Listing~\ref{lst:ue-specs}.
\begin{lstlisting}[float=ht,floatplacement=h,language=json,style=mystyle-json, 
caption={UE database example.}, 
label={lst:ue-specs}]
{
    "btdn0140013c": {
        "ue_hostname": "cicd-001",
        "ue_imsi": "001010000010666",
        "ue_ip_address": "10.112.1.53",
        "ue_location": "660/3",
        "ue_model": "NUC7i7DNKE",
        "ue_serial_number": "btdn0140013c"
    },
    "782452cbd384d676": {
        "ue_hostname": "cicd-002",
        "ue_imsi": "001010000012247",
        "ue_ip_address": "10.112.1.54",
        "ue_location": "660/4",
        "ue_model": "Raspberry Pi 5 Model B Rev 1.0",
        "ue_serial_number": "782452cbd384d676"
    },
    "mj06k2su": {
        "ue_hostname": "cicd-003",
        "ue_imsi": "001010000012252",
        "ue_ip_address": "10.112.1.52",
        "ue_location": "640/2",
        "ue_model": "10MUS30601",
        "ue_serial_number": "mj06k2su"
    }
}
\end{lstlisting}
For each \gls{ue}, this database---automatically updated by the \glspl{ue} via cron jobs---records information such as \gls{ue} hostname, \gls{imsi}, IP address, location, and model and serial number of the mini-PC controlling the Sierra Wireless \gls{5g} modem.
This automation enables continuous scalability of the system without requiring manual updates to the testing framework or \gls{ue} database.
Finally, once \glspl{ue} are connected to the \gls{gnb}, the specified traffic will start (for instance using tools such as iPerf), and results of the test will be collected and analyzed. In our tests, we deploy multiple iPerf servers to generate the traffic that the \glspl{ue} request during the test, and each instance is deployed as a pod.
Specific parameters of the test, such as server hostname and port, can be specified in the test files.
After the specified tests have been executed, performance metrics are extracted and stored in an InfluxDB database---together with test and protocol stack logs---and visualized via a Grafana dashboard, which also lets users compare the latest test results to historical data.
This, together with the automated update of the \gls{gnb} protocol stack is key for observing how new releases of the protocol stack software affect the network performance, enabling users to quickly spot degradations with respect to previous tests.

\begin{figure}[ht]
    \renewcommand{\thefigure}{8}
    \centering
    \includegraphics[width=\linewidth]{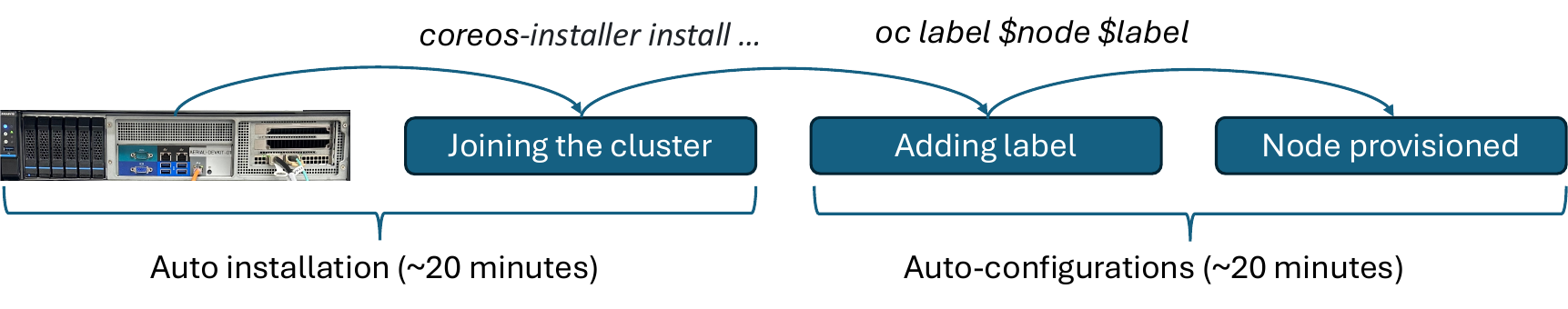}
    \caption{Steps required to provision a new node.}
    \label{fig:provisioning}
    \vspace{-.6cm}
\end{figure}

\begin{figure*}[t]
    \renewcommand{\thefigure}{9}
    \centering
    \begin{subfigure}[b]{0.32\textwidth}
        \centering
        \includegraphics[width=.9\linewidth]{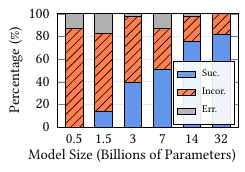}
        \caption{Success rate of deployments through LLM for different sizes of Qwen2.5 LLM.}
        \label{fig:success_LLM}
    \end{subfigure}
    \hfill
    \begin{subfigure}[b]{0.32\textwidth}
        \centering
        \includegraphics[width=.9\linewidth]{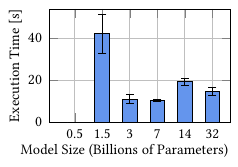}
        \caption{Runtime of LLM for successful deployments with different sizes of Qwen2.5 LLM.}
        \label{fig:runtime_LLM}
    \end{subfigure}
    \hfill
    \begin{subfigure}[b]{0.32\textwidth}
        \centering
        \includegraphics[width=.9\linewidth]{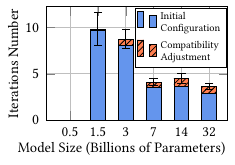}
        \caption{Number of iterations for successful deployments with different sizes of Qwen2.5 LLM.}
        \label{fig:iterations_LLM}
    \end{subfigure}
    \caption{Performance metrics for different sizes of Qwen2.5 LLM: (a) success rate, (b) runtime for successful deployments, and (c) number of iterations required.}
    \label{fig:llm_performance}
    \vspace{-.3cm}
\end{figure*}

\section{Experimental Evaluation}
\label{sec:experimental-evaluation}%

The section discusses several results that profile the performance and effectiveness of the proposed \framework approach. 
Specifically, in Section~\ref{sec:cluste-building} we review the time needed to onboard a new node in the cluster, comparing automation and manual solutions.
In Section~\ref{sec:llm_exp} we describe the \gls{llm} deployment and report related metrics.
In Section~\ref{sec:e2e-exp-eval}, we analyze the time needed to execute deployment and testing workflows using \framework, showing how a network based on microservices can be instantiated in a matter of seconds.
In Section~\ref{sec:performance-evaluation}, we provide performance metrics obtained from the deployed \glspl{gnb}.
In Section~\ref{sec:coexistence} we analyze coexistence between \gls{ran} and other generic workloads.
Finally, in Section~\ref{sec:resilience}, we provide insights on \framework's resiliency.

\vspace{-10pt}
\subsection{Onboarding Nodes on the \framework Cluster}
\label{sec:cluste-building}%
\framework is designed to ensure scalability. Even the extension of the cluster follows the same principle. To showcase the benefits of automated node creation and configuration, in this section we provide a comparison between adding a new \gls{ran} worker node manually and using \framework.
Manually deploying a NVIDIA ARC node requires performing several error-prone steps that need to be repeated on each compute node of the cluster just to set up the basic infrastructure. This usually requires additional effort to maintain the node with up-to-date drivers and software.
Using the profiles and configurations offered by \framework, extending the cluster with an already configured profile requires minimal human intervention, as shown in Figure~\ref{fig:provisioning} (i.e., it is only required to add the node to the cluster---using \texttt{coreos-installer}---and assign a label to it---using \texttt{oc label}), and guarantees a completely configured node ready to accept any targeted workload in approximately 40 minutes—without the same level of effort of manually configuring a node.
After their initial provisioning, nodes can be repurposed for different roles (e.g., edge vs \gls{ran} node) by modifying their labels. After this operation, nodes will reboot and auto-configure themselves as required.

\vspace{-10pt}
\subsection{LLM-Based Deployment}
\label{sec:llm_exp}
In this section, we evaluate our LLM-based deployment pipeline. To do so, we use Claude 3.7 Sonnet to generate a series of 33 different deployment prompts. Each prompt provides some specific requirements in \gls{naturallanguage} and is associated with the required element(s) in the deployed network. For example, if the prompt is \textit{``generate a 5G network based on GPU acceleration''}, the associated requirement is that the DU-low uses \gls{arc}. We evaluate the deployment pipeline by running each prompt 10 times, for different sizes of LLMs. 
The \gls{llm} is deployed on one of the control-plane nodes, which are equipped with a NVIDIA L40S GPU with $40$\:GB of VRAM.
Figure~\ref{fig:success_LLM} shows that the larger model is significantly better at providing satisfactory configurations, with Qwen2.5:32B reaching up to $80$\% correct responses, and with correctness consistently decreasing as we decrease the size of the model. Furthermore, Figure~\ref{fig:runtime_LLM} shows that this extra performance does not necessarily come at the cost of a larger execution time (measured from the time the user sends the prompt to the time the input to the Tekton pipeline is generated, see Fig.~\ref{fig:llm_diag}), with Qwen2.5:1.5B having the largest runtime despite its low success-rate.
We observe (see Figure~\ref{fig:iterations_LLM}) that this concurs with the high number of iterations in the initial phase (where we verify if values are missing, see Figure \ref{fig:llm_diag}): the model struggles to find any valid configuration until it times out. 
On the other hand, larger models have similar iteration counts due to their ability to output a correct configuration (in the sense that we almost always obtain a deployable network, but that network is not necessarily the one the user asked for, as per Figure~\ref{fig:iterations_LLM}).

\subsection{End-to-End Deployment and Testing Workflows}
\label{sec:e2e-exp-eval}

\begin{figure}[hb]   
  \centering
  \includegraphics[width=0.9\columnwidth]{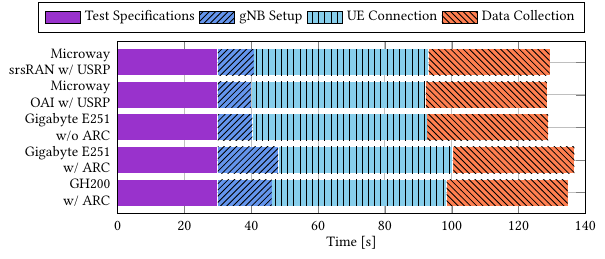}
  \caption{Average time duration of different tasks of the automated tests.}
  \label{fig:time-e2e}
  \vspace{-0.4cm}
\end{figure}

In this section, we show experimental results to validate the automated \gls{e2e} test workflows described in Section~\ref{sec:automated-workflows}, where we use Sierra Wireless \gls{5g} modems as \glspl{ue}.
We leave the extension of the automated tests to smartphone \glspl{ue} for future work, while we use these devices for manual and mobile tests.
Figure~\ref{fig:time-e2e} shows the time required to execute the different tasks of the automated testing pipeline discussed in Section~\ref{sec:automated-workflows}. 
Together, the following five bars show the average time required to perform a complete test with five different \gls{gnb} combinations: (i)~GH200 server with ARC; (ii)~Gigabyte E251 server with ARC; (iii)~Gigabyte E251 server without ARC; (iv)~Microway server with \gls{oai} and USRP; and (v)~Microway server with srsRAN and USRP.
The \textit{Test Specification} bar shows the time taken from when the user specifies which deployment and test configuration to use, to when the \gls{gnb} Flask API is called to start the test. 
The pull time, included in this value, is dependent on the size of the image and the network speed. Having a local and private registry inside OpenShift allows \framework to pull the image at the maximum speed allowed by the connection between the cluster and the \gls{nas}, that is up to $10$\:Gbps. Extra time is still required to the various layers of the image in the target node.
On the other hand, after the image has been pulled at least once on a node, it is thereafter kept in its cache registry, and subsequent pulls are in the order of a few milliseconds.
Therefore, by pre-loading the image in the internal cache on the node, the pull time becomes negligible even for the bigger images, such as cuBB (around $40$\:GB).
After the image has been deployed in a pod on the targeted node, the \gls{gnb} software needs to be instantiated (\emph{gNB setup} in Figure~\ref{fig:time-e2e}). 
As shown, the time to run the \gls{gnb} on the GH200 and Gigabyte servers is quite similar, at around $18$\:s.
This time accounts for the time to initialize libraries, GPU and \gls{nic} for the L1 acceleration.
On the other hand, since \revision{running} OAI without ARC on the Gigabyte server only one container is needed and the \gls{gnb} is monolithic, the start time is shorter, about $8$\:s.
As the \glspl{ru} \revision{are} always in idle state (i.e., in the on state but without actively transmitting or receiving) and initialized outside of OpenShift via M-plane-like functionalities before the \gls{gnb} pod deployment (see Section~\ref{sec:integration_details}), we do not include initialization (approximately $60$\:s for a Foxconn \gls{ru}) and reconfiguration times in the instantiation time of \gls{cu}/\gls{du}.
This is different from the \glspl{usrp} deployment, where initialization times are negligible.
Indeed, the time taken to start the \gls{gnb} and connect it to the USRP for both srsRAN and OAI is around $10$\:s.
Finally, the last two bars show the time taken for the \gls{ue} (Sierra Wireless \gls{5g} modem in our case) to connect to the gNB, and the time taken to collect the results and send them to the data collector pod respectively.
It is worth mentioning that we do not report the duration of the data transmissions between \gls{gnb} and \gls{ue}, as it depends on the value in the test specifications (e.g., run an iPerf test for $60$\:s, see Listing~\ref{lst:traffic test-example}).
Finally, multiple tests can be executed either as independent tests (e.g., one for iPerf with TCP and one for iPerf3 in UDP) where the \gls{gnb} is restarted prior to the execution of the next test, or consecutively without redeploying the pod.

\begin{figure}[t]
    \centering
    \includegraphics[trim={0 4.5pt 0 0},clip,width=.95\linewidth]{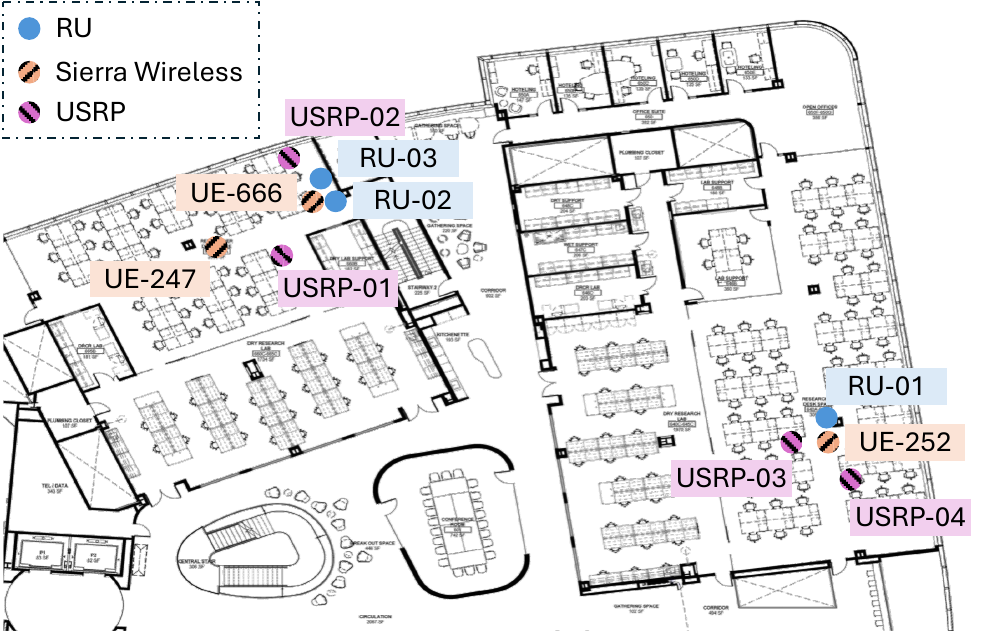}
    \caption{Location of RUs, Sierra Wireless UEs, and USRPs used in our tests.}
    \label{fig:testbed}
    \vspace{-.6cm}
\end{figure}

\vspace{-10pt}
\subsection{Performance Evaluation of \ran Functions}
\label{sec:performance-evaluation}%
We show results on the performance of the \ran stacks obtained using the \framework framework. In more detail, in Section~\ref{sec:perforance-evaluation-sierra}, we show metrics obtained through the automated testing of the Sierra Wireless modems deployed in static location. In Section~\ref{sec:performance-arc}, we provide a comparison of the raw performance of the L1 accelerator from NVIDIA between the original NVIDIA ARC deployment and \framework's.

For over-the-air tests, the locations of the \glspl{ru} and \glspl{ue} used in this test within our laboratory are shown in Figure~\ref{fig:testbed}. Specifically, we used \gls{ue}~252 and Foxconn RU-03. For the experiments with the USRPs, the USRP antennas are chosen to be at the same distance from the \gls{ue} as that between \gls{ue} and \gls{ru} in the experiments with the Foxconn radio.

\subsubsection{Performance of the Automated Testing Workflow}
\label{sec:perforance-evaluation-sierra}%
\begin{figure}[t]
  \centering
  \includegraphics[width=\columnwidth]{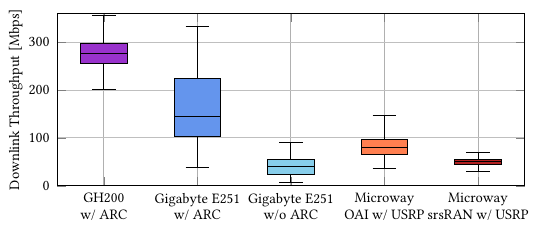}
  \caption{Bar plot of achieved TCP downlink datarate of Sierra 5G modem with five different configurations.}
  \label{fig:dl-tcp-box-plot}
        \vspace{-.3cm}
\end{figure}

\begin{figure}[t]
  \centering
  \includegraphics[width=\columnwidth]{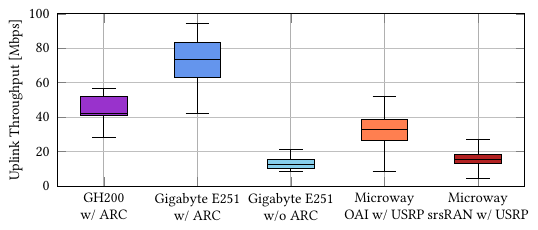}
  \caption{Bar plot of achieved TCP uplink datarate of Sierra 5G modem with five different configurations.}
  \label{fig:ul-tcp-box-plot}
      \vspace{-.6cm}
\end{figure}

Figures~\ref{fig:dl-tcp-box-plot} and~\ref{fig:ul-tcp-box-plot} show the downlink and uplink TCP throughput achieved during the specific \gls{e2e} experiments performed through the workflows described in Section~\ref{sec:automated-workflows} to test five different combinations of \glspl{gnb} with a single Sierra Wireless as the \gls{ue}. For this specific test, we select TCP as a way to measure not only raw performance but also the realistic goodput reaching the considered \gls{ue}.

We notice that in the downlink direction, the \gls{gnb} deployed with ARC on the GH200 server achieves an average throughput of $275$\:Mbps, which surpasses the throughput of both the ARC \gls{gnb} deployed on the Gigabyte server and that of the deployments without ARC.
These experiments also show that \gls{oai} achieves slightly better performance than srsRAN when they are both deployed on the Microway server and use USRP \glspl{sdr} over a $60$\:MHz bandwidth.
Uplink performance with the Sierra Wireless modem, instead, reaches more than $75$\:Mbps when using the ARC setup on the Gygabite server, which outperforms the four other \gls{gnb} configurations.

\subsubsection{Performance Comparison of NVIDIA ARC}
\label{sec:performance-arc}
\begin{figure}[t]
    \centering
    \includegraphics[width=\columnwidth]{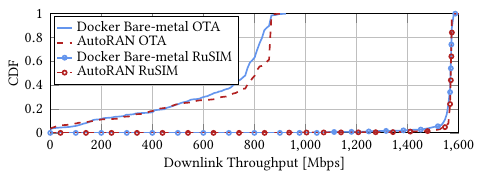}
    \caption{Downlink comparison of OTA and RuSIM transmissions between using \framework and Docker on bare metal on the Grace Hopper.}
    \label{fig:arc_comparison_4cdfs}
    \vspace{-.6cm}
\end{figure}

In this section, we focus on comparing, for the first time, an OpenShift-based deployment of NVIDIA ARC with its counterpart based on bare-metal and Docker virtualization.
In the following, the performance we report is obtained from the cuBB L1 accelerator logs (and not from the \gls{ue}), since we are interested in comparing the raw throughput being pushed by the accelerator, also considering the partitioned \gls{nic} and the extra levels of virtualization.
We run a comparison using \gls{ota} transmissions with a smartphone and the RuSIM emulator. Results of both experiments are shown in Figure~\ref{fig:arc_comparison_4cdfs}. For the former configuration, we generate downlink traffic toward a single Samsung S23 phone \gls{ue} in \revision{an} \gls{ota} transmission using a Foxconn \gls{ru} in the N78 band configured at $100$\:MHz bandwidth with DDDDDDSUUU as \gls{tdd} pattern. We test iperf3 connectivity between the phone and a server pushing $1$\:Gbps to the phone itself for a duration of $60$\:s. The \gls{cdf} is calculated as average of 5 experiments. Note that, compared to the Sierra Wireless boards used in the previous tests, this device supports 4 layers in downlink, resulting in higher achievable throughput. The results, indeed, show that the throughput in downlink reaches up to $820$\:Mbps, with similar performance for the \framework and bare metal configuration. 

To exclude any variability related to the radio environment, we also perform an emulated comparison using RuSIM, one of the Keysight tools available in the Northeastern University OTIC~\cite{gemmi-otic}. RuSIM is a \gls{ru} emulator for 7.2 split, mirroring all parameters that a real \gls{ru} should expose. In addition to this, it allows simulation of users and of wireless channel. It is usually used jointly with other Keysight tools, including CoreSIM (to emulate core network and traffic generation) and Air Mosaic for managing and orchestrating tests. 
\framework provides a flexible testing platform and the integration of RuSIM only requires the addition of a second \gls{sriov} interface to communicate with CoreSIM, since it is external to the cluster. This shows the flexibility of the \framework virtualization capability, since the same physical interface is used twice, for fronthaul and backhaul, using two different virtual functions.
We test performance of 15 simulated users, which are able to achieve an aggregate performance of around $1600$\:Mbps using all slots ($1600$) every second available over $100$\:MHz with a DDDDDDDSUU \gls{tdd} pattern. 

Our results show that the added virtualization and integration into \framework does not lead to relevant performance difference in raw throughput when deploying NVIDIA ARC \glspl{gnb}.
It is worth mentioning that the cuBB logs show some delayed slots from time to time in the OpenShift-based implementation (probably due to the missing optimized kernel), but this situation does not seem to affect performance.
Therefore, the \framework approach shows how it is possible to apply infrastructure sharing to the \gls{ran} world as in the cloud computing one without significant performance degradations.

\begin{figure}[h]
  \centering
  \includegraphics[width=0.9\columnwidth]{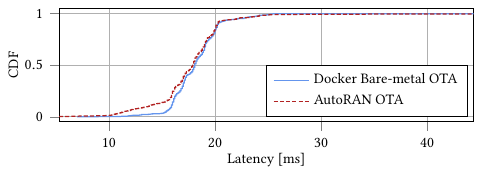}
  \caption{CDF plot of achieved ICMP latency of a \gls{cots} \gls{ue}.}
  \label{fig:ping-cdf-plot}
\end{figure}

As at the time of this writing, open-source \gls{ran} stacks are not optimized for \gls{urllc} use cases, we evaluate latency by exchanging ICMP packets between a Samsung Galaxy S23 \gls{cots} \gls{ue} and core network on \framework as well as on the bare-metal NVIDIA ARC deployment. Results for this evaluation are shown in Figure~\ref{fig:ping-cdf-plot}. We notice that the performance between the original NVIDIA ARC and the \framework deployments are similar, with small variations in latency related to non-deterministic channel conditions and an average ICMP \gls{rtt} of $18$\:ms.

\subsection{Coexistence with Different Workloads}
\label{sec:coexistence}
\begin{figure}[ht]
    \centering
    \vspace{-0.3cm}
    \includegraphics[width=0.95\linewidth]{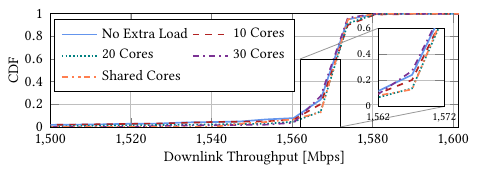}
    \caption{Performance comparison when applying load on either $N$ isolated and shared cores using RuSIM at full capacity.}
    \label{fig:coexistence-rusim}
\end{figure}

In this section, we test whether generic workloads on a \gls{ran} oriented node has any effect on the \gls{ran} itself. We want to evaluate this coexistence capabilities for various reasons, including building of containers on the targeted node and any other workloads that the OpenShift scheduler allocates on that node itself on the shared CPUs.
We show how running increased CPU-based consumption for a long period of time does not affect the throughput of the L1. We use the RuSIM emulation in Figure~\ref{fig:coexistence-rusim}. 
We simulate high demanding workloads using the Linux tool \texttt{stress} on either $10,20,30$ isolated CPUs and on the generally available shared CPUs of the system.

\begin{figure}[ht]
    \centering
    \includegraphics[width=0.8\linewidth]{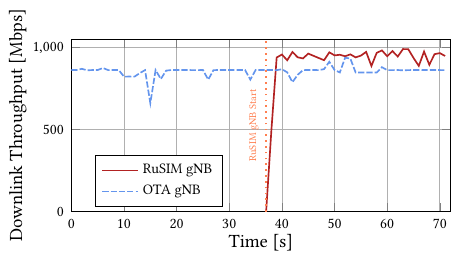}
    \caption{\revision{Performance comparison with concurrent \gls{ran} workloads using NVIDIA ARC.}}
    \label{fig:coexistence-ran}
    \vspace{-0.6cm}
\end{figure}

\revision{In Figure~\ref{fig:coexistence-ran}, we show the coexistence of multiple \glspl{ran} on the same node using the \gls{sriov} and \gls{mig} partitioning technologies on the Grace Hopper node. In this experiment, we consider two cells at $100$\:MHz: one based on RuSIM, and the other \gls{ota} with one \gls{cots} \gls{ue} attached. 
The L1s of both \glspl{ran} are completely independent in CPUs, memory, and GPU. We show that the throughput of the first active cell is not impacted by the activation of the second cell, despite sharing the same GPU.
Latency is not affected by the coexistence of multiple \glspl{gnb}, with the same average results as shown in Figure~\ref{fig:ping-cdf-plot}}.
This \revision{set} of experiments shows how \framework does not introduce relevant delays or performance degradation into the \gls{gnb} stack while using an additional layer of virtualization and exposing a new set of features that streamlines and simplifies 5G lifecycle management.

\subsection{\framework Resilience}%
\label{sec:resilience}%
The OpenShift container orchestration platform \framework is based upon---which is, in turn, based on Kubernetes---natively provides high-availability mechanisms to compensate for node failures by ``evicting'' (i.e., moving) workloads from unresponsive to functioning cluster nodes. 
As mentioned in Section~\ref{sec:e2e-telco-cloud}, however, \gls{ran} workloads typically require specific hardware functionalities to be available on the compute nodes. Therefore, they can only move across nodes where these are available (in practice, across nodes part of the same \gls{mcp}), while an \framework-orchestrated redeployment is needed across nodes with different \glspl{mcp}. 
The eviction happens automatically after a timeout that starts when OpenShift detects the node to be unavailable. The duration of the timeout is configurable between $0$\:s (the pod gets evicted as soon as the node appears unhealthy) and infinite (the pod never gets evicted), with a default of 5 minutes. Further analysis of this parameter is left for future work.
Additionally, due to limitations of the protocol stack implementations we leverage, a new instantiation of the pipeline is required to re-start the \gls{gnb} once the workload is re-allocated. 

\begin{figure}[ht]
  \centering
  \vspace{-.5cm}
  \includegraphics[width=\columnwidth]{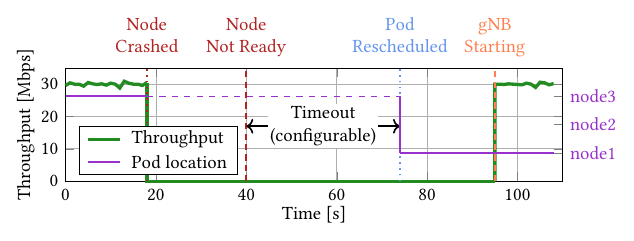}
  \caption{Workflow of \framework when a node is detected to be unavailable.}
  \label{fig:resiliency}
\end{figure}

\revision{Figure~\ref{fig:resiliency} shows an experiment where the OpenShift scheduler detects a failed node (Microway) in around $40$\:s. Once this happens, OpenShift enforces a timeout (set at $30$\:s), after which the pod is up again and ready to start a new transmission or deployment in a new node of the same \gls{mcp}. The \gls{ue}, generating around $30$\:Mbps of traffic, gets disconnected, but it reconnects and is able to generate traffic again around $30$\:s after the \gls{gnb} has been redeployed, similarly to what described in Figure~\ref{fig:time-e2e}.}
Note that, during this process, \glspl{ue} can detect a \gls{rlf} and connect to a different cell, or directly reconnect to the new \gls{gnb}, depending on the duration of the eviction timer.
\revision{In the specific case of a 7.2 \gls{gnb}, the re-deployment of a \gls{du}} triggers an \gls{ru} reset to accommodate the new \gls{du} MAC address (alternatively, the \gls{du} MAC can be spoofed, although this approach would not be scalable without resorting to additional middleware components, as discussed in~\cite{foukas2025ranbooster}). 

\vspace{-0.2cm}
\section{Conclusions and Future Work}
\label{sec:conclusions}%
Starting from the Open \gls{ran} principles of disaggregation and standardization, we proposed \framework, an open automation framework, which leverages cloud computing and virtualization techniques to seamlessly deploy, reconfigure, and test heterogeneous private \gls{5g} \glspl{ran} and related components. 
We showcased the building blocks of \framework, highlighting its advantages with respect to traditional deployment techniques, which are often monolithic and designed for highly experienced users. 
We experimentally evaluated the capabilities of \framework, showing how the virtualization and automation functionalities that it offers introduce enhanced flexibility and seamless deployment and testing capabilities that take in input simple high-level intents expressed in natural language, instead of complex and detailed configurations. 
In future works, we will extend the same approach to manage additional \glspl{nic} and hardware combinations 
(including \gls{ru} configuration through M-Plane),
as well as the automatic deployment of auxiliary components to the \gls{ran}.
\vspace{-0.4cm}

\section*{Acknowledgments}
\label{sec:Acknowledgments}
The authors would like to express their gratitude to Jing Xu, Anupa Kelkar, and Chris Dick from NVIDIA Corporation for their feedback on NVIDIA ARC.
The authors disclose that AI tools (ChatGPT, Claude) have been used to perform minor edits and reformulations of the text.
\vspace{-0.4cm}

\bibliographystyle{IEEEtran}
\bibliography{biblio}

\vspace{-1cm}

\begin{IEEEbiography}[{\includegraphics[width=1in,height=1.25in,clip,keepaspectratio]{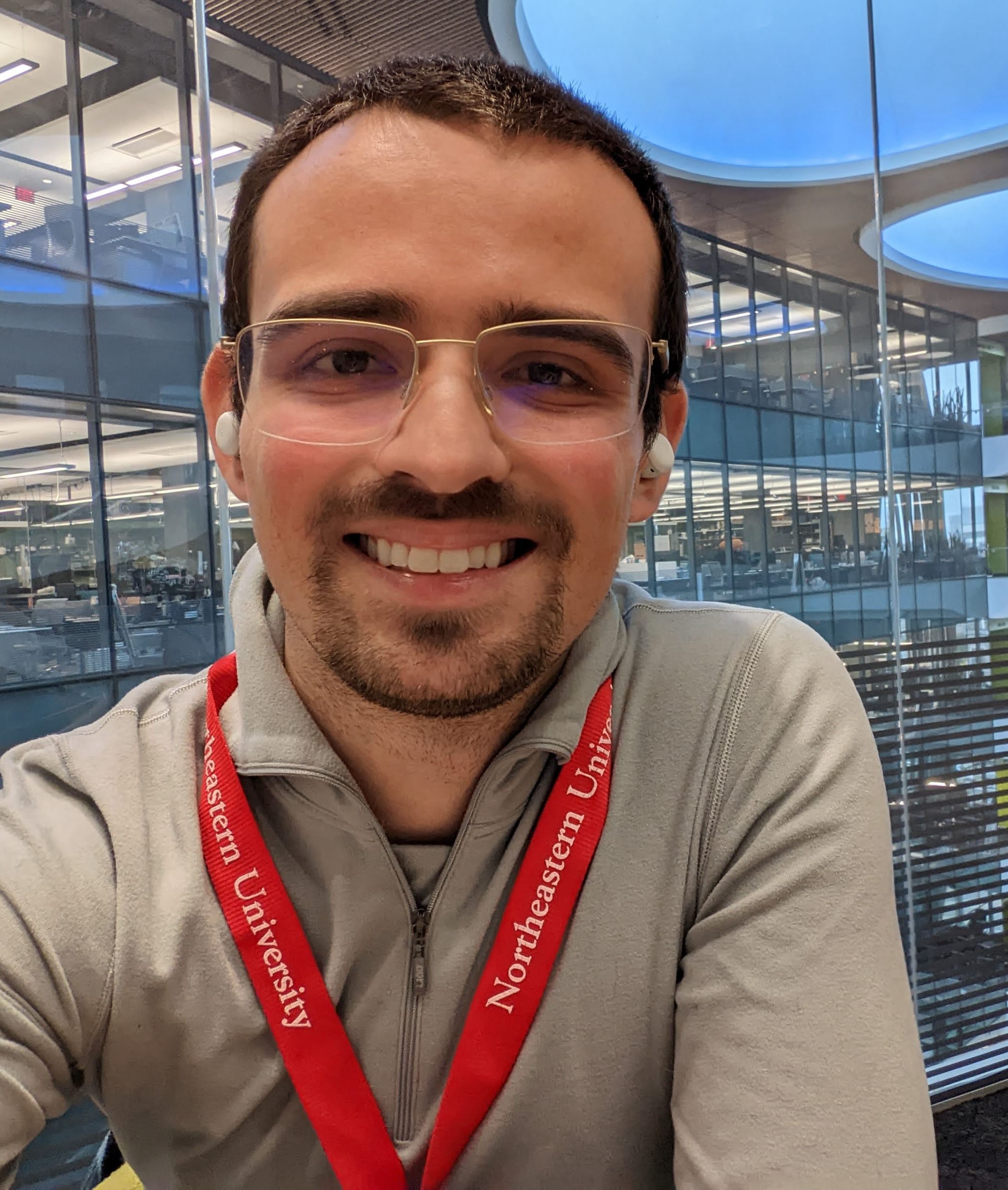}}]{Stefano Maxenti} is a Ph.D. Candidate in Computer Engineering at the Institute for the Wireless Internet of Things (WIoT) at Northeastern University, under Prof. Tommaso Melodia. He received a B.Sc. in Engineering of Computing Systems in 2020 and a M.Sc. in Telecommunication Engineering in 2023 from Politecnico di Milano, Italy. His research is linked with AI applications for wireless communications and orchestration, integration, and automation of O-RAN networks.
\end{IEEEbiography}

\vspace{-1cm}

\begin{IEEEbiography}[{\includegraphics[width=1in,height=1.25in,clip,keepaspectratio]{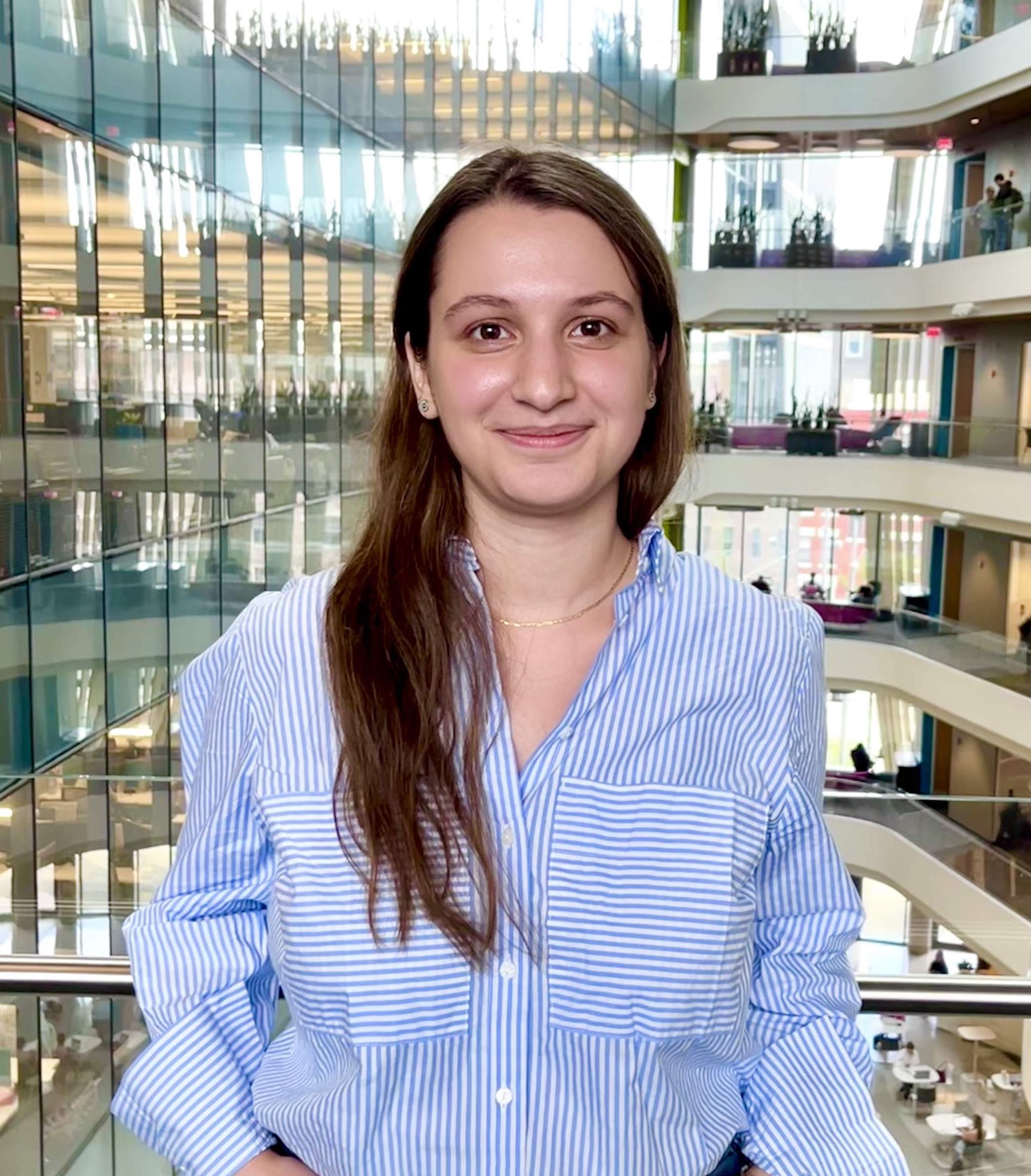}}]{Ravis Shirkhani} is a Ph.D. Candidate in Computer Engineering at the Institute for the Wireless Internet of Things (WIoT) at Northeastern University, under Prof. Tommaso Melodia. She received a B.Sc. in Electrical Engineering (Communication Systems and Networks) in 2023 from Sharif University of Technology, Iran. Her research focuses on automation of O-RAN networks, power consumption across O-RAN components, and exploring optimization approaches for network energy efficiency.
\end{IEEEbiography}

\vspace{-1cm}

\begin{IEEEbiography}[{\includegraphics[width=1in,height=1.25in,clip,keepaspectratio]{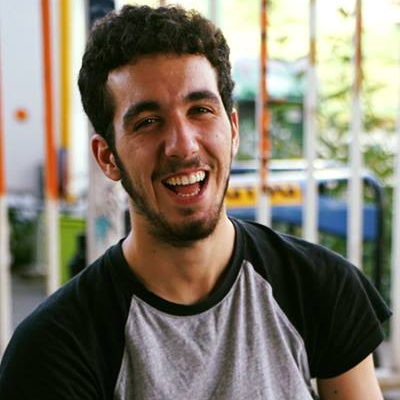}}]{Maxime Elkael} is a Research Scientist at the Institute for the Wireless Internet of Things at Northeastern University. He received his Ph.D in Computer Science from Institut Polytechnique De Paris/Telecom SudParis in 2023. His research interest lies at the intersection of optimization theory, artificial intelligence and graph theory applied to next generation wireless networks, especially Open RAN networks.
\end{IEEEbiography}

\vspace{-1cm}

\begin{IEEEbiography}[{\includegraphics[width=1in,height=1.25in,keepaspectratio]{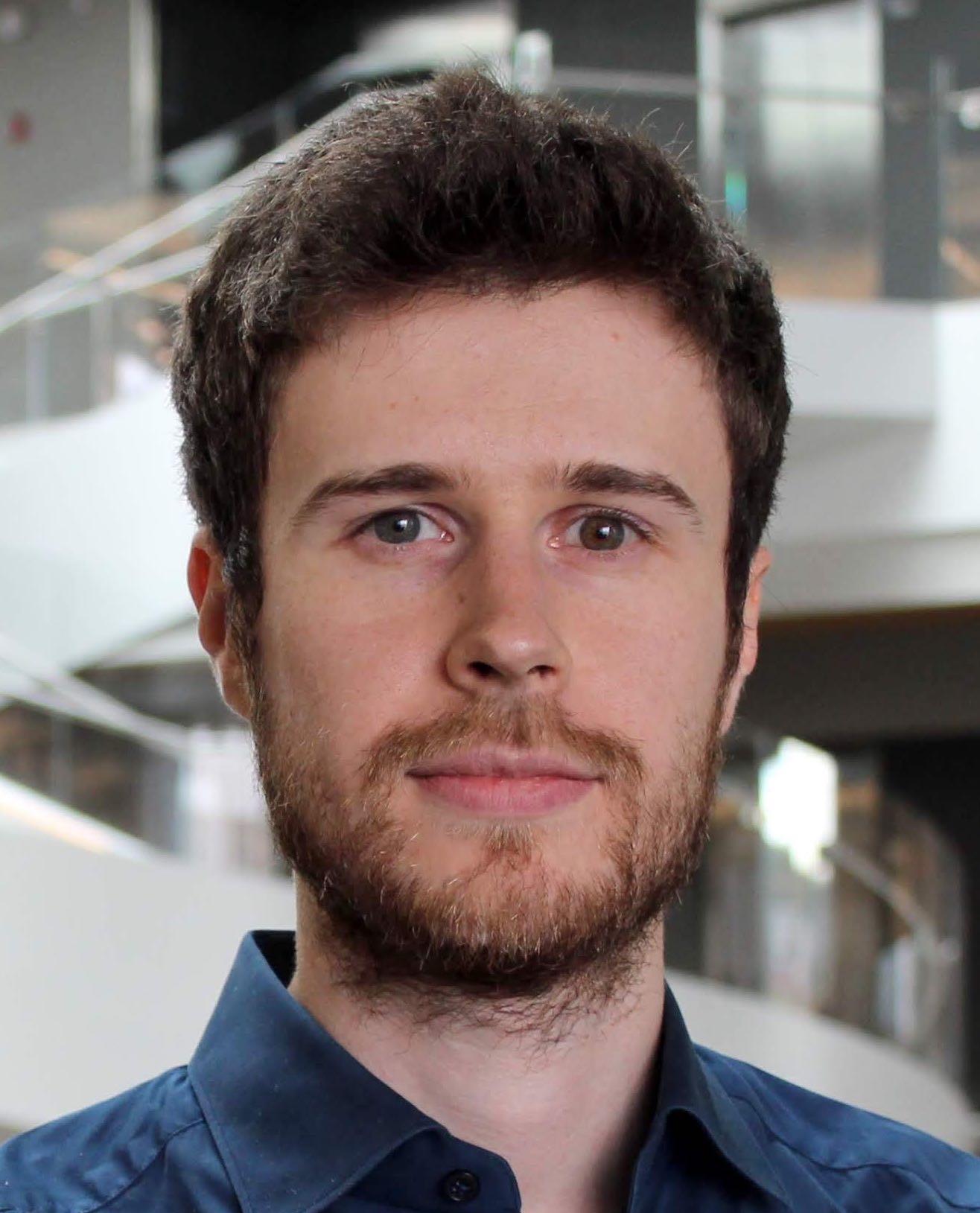}}]{Leonardo Bonati} is an Associate Research Scientist at the Institute for the Wireless Internet of Things, Northeastern University, Boston, MA. He received a Ph.D. degree in Computer Engineering from Northeastern University in 2022. His main research focuses on softwarized approaches for the Open Radio Access Network (RAN) of the next generation of cellular networks, on O-RAN-managed networks, and on network automation, orchestration, and virtualization. He was awarded the 2024 Mario Gerla Award for Research in Computer Science. Leonardo served as TPC co-chair for the IEEE DTwin 2025 workshop, as co-chair for the track on Testbeds, Experimentation and Datasets for Communications and Networking of IEEE CCNC 2025, and as guest editor of the special issue of Elsevier Computer Networks on Advances in Experimental Wireless Platforms and Systems.
\end{IEEEbiography}

\vspace{-40pt}

\begin{IEEEbiography}[{\includegraphics[width=1in,height=1.25in,clip,keepaspectratio]{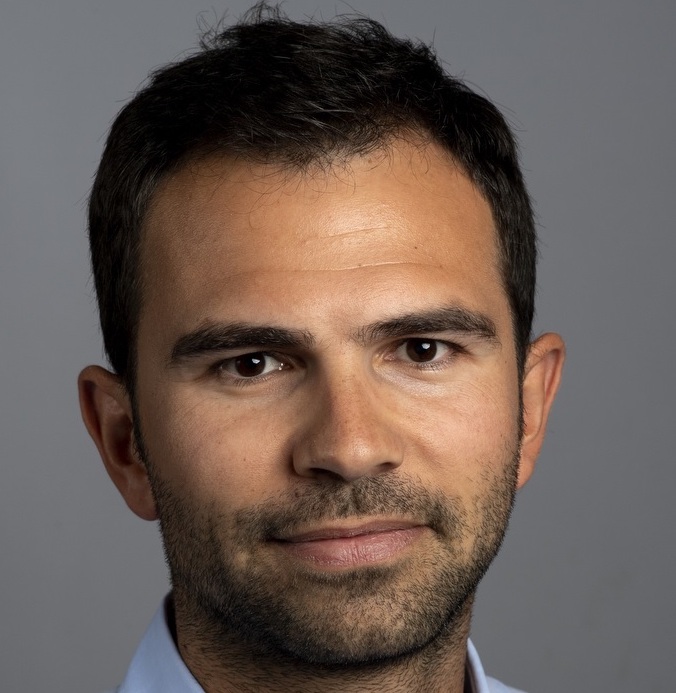}}]{Salvatore D'Oro} is the CTO and co-founder of zTouch Networks, a company focused on the development of zero-touch automation solutions for O-RAN systems. He is also a Research Associate Professor at Northeastern University. He received his Ph.D. degree from the University of Catania and is an area editor of Elsevier Computer Communications journal. He serves on the TPC of IEEE INFOCOM, IEEE CCNC \& ICC and IFIP Networking. He is one of the contributors to OpenRAN Gym, the first open-source research platform for AI/ML applications in the Open RAN. His research interests include optimization, AI \& network slicing for NextG Open RANs.
\end{IEEEbiography}
\vspace{-3\baselineskip}

\begin{IEEEbiography}[{\includegraphics[width=1in,height=1.25in,clip,keepaspectratio]{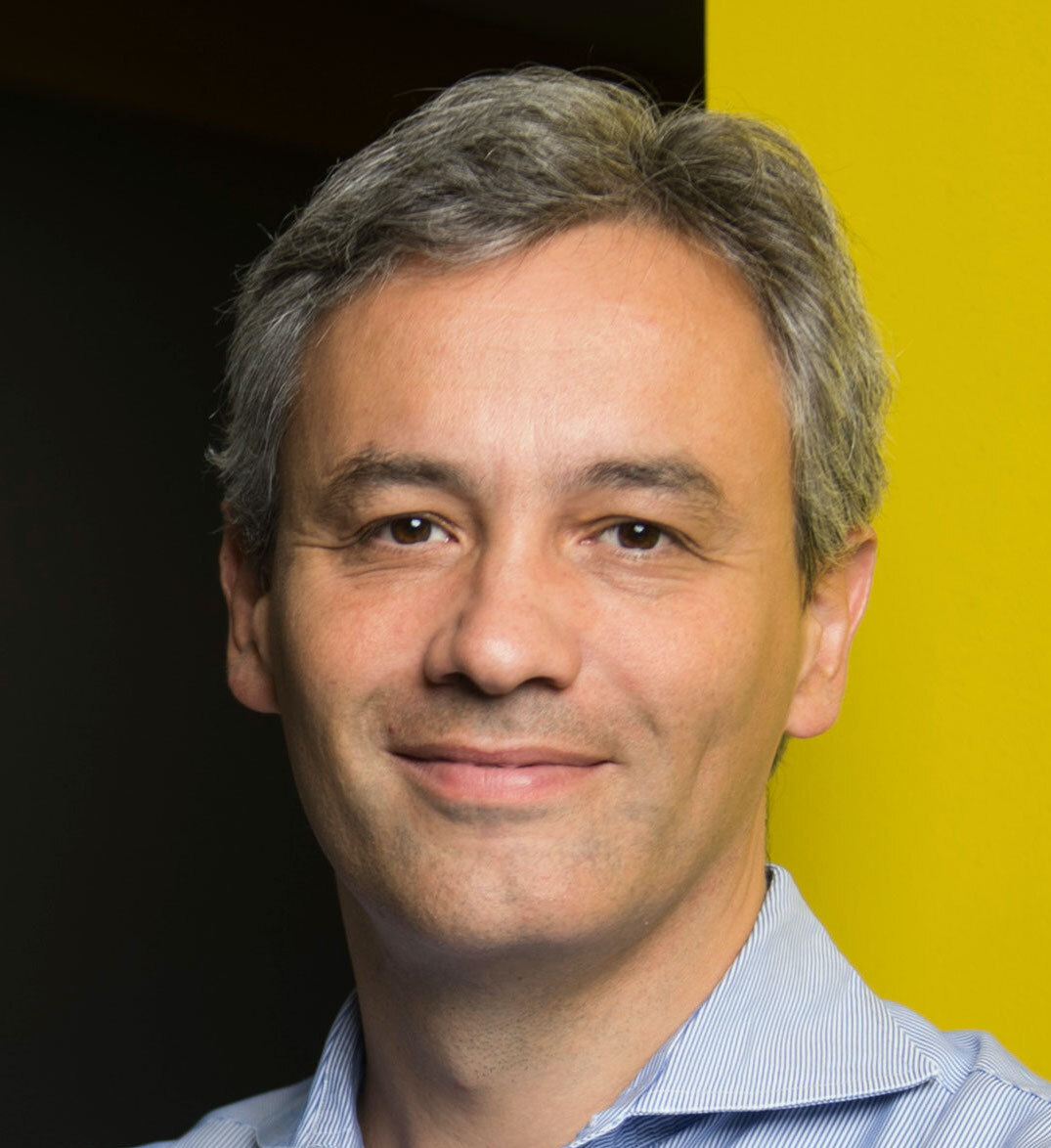}}]{Tommaso Melodia}
is the William Lincoln Smith Chair Professor with the Department of Electrical and Computer Engineering at Northeastern University in Boston. He is also the Founding Director of the Institute for the Wireless Internet of Things and the Director of Research for the PAWR Project Office. He received his Ph.D. in Electrical and Computer Engineering from the Georgia Institute of Technology in 2007. He is a recipient of the National Science Foundation CAREER award. Prof. Melodia has served as Associate Editor of IEEE Transactions on Wireless Communications, IEEE Transactions on Mobile Computing, Elsevier Computer Networks, among others. 
He has served as Technical Program Committee Chair for IEEE INFOCOM 2018, General Chair for IEEE SECON 2019, ACM Nanocom 2019, and ACM WUWnet 2014. 
Prof. Melodia is the Director of Research for the Platforms for Advanced Wireless Research (PAWR) Project Office, a \$100M public-private partnership to establish four city-scale platforms for wireless research to advance the US wireless ecosystem in years to come. Prof. Melodia's research on modeling, optimization, and experimental evaluation of Internet-of-Things and wireless networked systems has been funded by the National Science Foundation, the Air Force Research Laboratory the Office of Naval Research, DARPA, and the Army Research Laboratory. Prof. Melodia is a Fellow of the IEEE and a Distinguished Member of the ACM.
\end{IEEEbiography}

\begin{IEEEbiography}[{\includegraphics[width=1in,height=1.25in,clip,keepaspectratio]{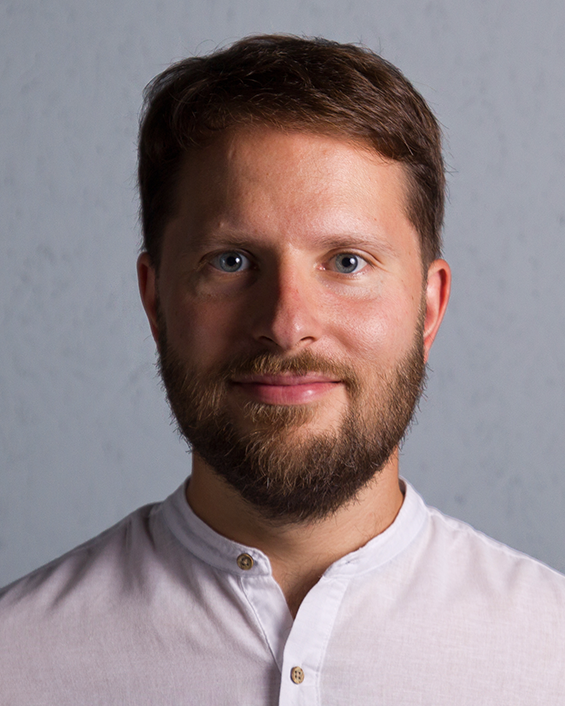}}]{Michele Polese} is a Research Assistant Professor at the Institute for the Wireless Internet of Things, Northeastern University, Boston, since October 2023. He received his Ph.D. at the Department of Information Engineering of the University of Padova in 2020. He then joined Northeastern University as a research scientist and part-time lecturer in 2020. 
During his Ph.D., he visited New York University (NYU), AT\&T Labs in Bedminster, NJ, and Northeastern University.
His research interests are in the analysis and development of protocols and architectures for future generations of cellular networks (\gls{5g} and beyond), in particular for millimeter-wave and terahertz networks, spectrum sharing and passive/active user coexistence, open RAN development, and the performance evaluation of end-to-end, complex networks. He has contributed to O-RAN technical specifications and submitted responses to multiple FCC and NTIA notice of inquiry and requests for comments, and is a member of the Committee on Radio Frequency Allocations of the American Meteorological Society (2022-2024). He is PI and co-PI in research projects on 6G funded by the NTIA, the O-RAN ALLIANCE, U.S. NSF, OUSD, and MassTech Collaborative, and was awarded with several best paper awards and the 2022 Mario Gerla Award for Research in Computer Science. Michele is serving as TPC co-chair for WNS3 2021-2022, as an Associate Technical Editor for the IEEE Communications Magazine, as a Guest Editor in an IEEE JSAC Special Issue on Open RAN, and has organized the Open 5G Forum in Fall 2021 and the NextGenRAN workshop at Globecom 2022.
\end{IEEEbiography}


\balance

\end{document}